\documentclass[fleqn,usenatbib]{mnras}

\usepackage{newtxtext,newtxmath}

\usepackage[T1]{fontenc}

\DeclareRobustCommand{\VAN}[3]{#2}
\let\VANthebibliography\thebibliography
\def\thebibliography{\DeclareRobustCommand{\VAN}[3]{##3}\VANthebibliography}

\usepackage{graphicx}
\usepackage{amsmath}
\usepackage{float}

\newcommand{\COtrap}{$\mathcal{T}_{\rm{CO}}~$}

\title[Locked In Ice]{Locked In Ice: how Pebble Drift and Volatile Entrapment can Significantly Impact Carbon and Oxygen Ratios in Evolving Protoplanetary Discs}

\author[Joe Williams et al.]{
Joe Williams,$^{1}$\thanks{E-mail: jw1436@exeter.ac.uk}
Sebastiaan Krijt,$^{1}$
Bertram Bitsch,$^{2}$
Adrien Houge,$^{3}$
and Jennifer Bergner$^{4}$
\\
$^{1}$School of Physics and Astronomy, University of Exeter, Stocker Road, Exeter EX4 4QL, UK\\
$^{2}$Department of Physics, University College Cork, Cork, T12 R229, Ireland\\
$^{3}$Center for Star and Planet Formation, GLOBE Institute, University of Copenhagen, Øster Voldgade 5-7, DK-1350 Copenhagen, Denmark\\
$^{4}$UC Berkeley Department of Chemistry, Berkeley, CA 94720, USA
}

\date{Accepted 2025 October 16. Received 2025 October 3; in original form 2025 July 25}

\pubyear{2025}

\begin{document}
\label{firstpage}
\pagerange{\pageref{firstpage}--\pageref{lastpage}}
\maketitle

\begin{abstract}
The complex interplay between the growth, drift, and sublimation of ice-covered pebbles can strongly influence the volatile distribution and evolution of disc composition, and therefore impact the composition of forming planets. Classic pebble drift models treat volatile species individually as sublimating at their respective snowlines, although observations from the James Webb Space Telescope (JWST) suggest that ices are likely mixed; laboratory studies suggest ice mixtures can exhibit more complex sublimation behaviours, remaining trapped beyond their nominal sublimation temperatures. We present the first model that couples pebble growth and drift with CO entrapment inside water ice - preventing a fraction (up to $\sim$60\%) of the CO from sublimating at its snowline, instead desorbing via volcanic desorption at the water crystallisation front, at 130K. Our models show that CO entrapment will significantly impact the carbon and oxygen distributions, enhancing the gas-phase C/O and C/H inside the water snowline by up to a factor of 10 over 1 Myr and a factor of a few around the CO$_2$ snowline; O/H is also increased around the CO$_2$ snowline, but is water-dominated in the inner disc. Entrapment therefore provides a means of introducing more carbon to the inner disc whilst retaining a large amount of water. We discuss connections to planet formation, noting that CO entrapment can increase the gas-phase heavy element content around the water snowline by up to 150\%. We also consider links to JWST observations and highlight the importance of entrapment for pebble drift models to accurately model disc composition.
\end{abstract}


\section{Introduction}

Characterising exoplanets has been a subject of study for many years, and the advent of JWST has made detections significantly more sensitive to atmospheric composition \citep[e.g.][see \citealt{kempton&knutson2024_jwst_exo_atmo} for a review]{zieba2022_k218b, greene2023_jwst_earth_exo, madhusudhan2023_carbon_hycean, kempton2023_jwst_metal_atmo}. The content of their atmospheres has long been connected to their formation pathways \citep[e.g.][]{oberg2011_snowlines, madhusudhan2014_hot_jupiter_chem_form, mordasini2016_jupiter_atmo_form, eistrup2016_volatile_evolution, eistrup2018_evolving_C/O, booth2017_giant_enrich, booth&ilee2019_pebble_chem, cridland2019_formation_and_chem, schneider&bitsch2021b_atmospheres, johansen2021_solar_system, molliere2022_exoplanet_atmo, wang2023_atmos_recycle}, although the link between C/O in protoplanetary discs and exoplanet atmosphere may not be as straightforward as first thought \citep[][and references therein]{molliere2022_exoplanet_atmo,  bitsch2022_wasp77Ab, bergin2024_c/o_wide_exoplanets, penzlin2024_formation_histories}. One major complication is that the disc composition is closely linked to the inward drift of ice-covered pebbles, which release a wealth of chemical species as their migrate to the inner disc.

The evolving distribution of volatiles in protoplanetary discs has been studied extensively with pebble drift models \citep[e.g.][]{piso2015_desorption_dist, booth2017_giant_enrich, stammler2017_CO, booth&ilee2019_pebble_chem, krijt2020_2D_CO_depletion, kalyaan2021, schneider&bitsch2021a_chemcomp, mah2023_vlms_C/O, mah2024_mind_the_gap} and has been consistently linked to the distributions observed in discs using ALMA \citep[e.g.][]{zhang2019_systematic_variations, zhang2020_excess_carbon, zhang2021_MAPS, bosman2023_im_lup, jiang2023_giant_footprint, williams&krijt2025_mass_constraint} and JWST \citep[e.g.][]{banzatti2020_pebble_drift, banzatti2023_excess_water, long2025_30myr_disc, gasman2025_disc_structure_volatiles, vlasblom2025_co2_rich_drift, sellek2025_co2_drift, houge2025_smuggled_water, arabhavi2025_minds_VLMS}. 

In addition to linking the gas-phase chemical reservoir of the inner region ($\lesssim5~\rm{au}$) of discs to pebble drift, JWST has also been used to study the chemical inventory of the ice mantle of pebbles in the outer regions, prior to sublimation, through ice absorption spectra \citep[e.g.][]{mcclure2023_jwst_ice_age, sturm2023b_HH48NE_ices, sturm2023c_ice_age, bergner2024_jwst_ice, brunken2024_envelope_ices_jwst, ballering2025_water_ice_jwst}. These ice absorption spectra reveal key insights into the spatial distribution and microphysical structure of ices on pebbles, located a few scale-heights away from the midplane \citep{sturm2023a_HH48NE_geometry, sturm2023b_HH48NE_ices, sturm2024_hh48ne_ice_distribution}, suggesting that their structure may be mixed rather than pure, distinct ice layers \citep{sturm2023c_ice_age, bergner2024_jwst_ice}.

 Radiative transfer models from \citet{bergner2024_jwst_ice} that best match the ice absorption spectra for the HH 48 NE disc feature CO mixed inside CO$_2$ ice as well as CO and CO$_2$ ice mixed inside water ice. Mixing these ices together prevents all of the CO budget, for example, from sublimating at the CO snowline; instead, a significant fraction will non-thermally desorb via `volcanic desorption' as the water is transformed from amorphous to crystalline form at 130K \citep{smith1997_molecular_volcano, bar-nun2007_N2_Ar_CO_trap, burke&brown2010_ice_desorption}, with a smaller portion of the budget co-desorbing with CO$_2$. Mixing these ices inside each other is often referred to as `trapping' or `volatile entrapment' \citep[e.g.][]{collings2003_CO_trapping}.

The entrapment of volatiles inside water ice has been studied extensively in laboratories for a long time \citep[see e.g.][and references therein]{bar-nun1985_ice_trapping, bar-nun1988_amorphous_water_trap, collings2003_CO_trapping, collings2004_thermal_desorb, collings2004_water_co2_entrapment, burke&brown2010_ice_desorption, fayolle2011_co2_trap_in_ice, simon2019_co_trap_in_co2, simon2023_comet_entrapment, kipfer2024_N2_trapping, zhou_competitive_2024, pesciotta2024_matrix_thick}. It has, however, been neglected in models of volatiles in protoplanetary discs \citep{schneeberger2023_protonebula_ices} and the impact of volatile trapping on volatile distributions and its implications for planet formation have not been extensively studied.

Recent work by \citet{ligterink2024_trapping} examined the impact of trapping CO and CO$_2$ in water ice, and CO in CO$_2$ ice on C/O ratios in a static disc and compared it to results from \citet{oberg2011_snowlines}. They then considered the importance of trapping in the context of planet formation and characterising the atmosphere of forming planets. They found that the solid-phase C/O ratio was changed from $\sim$0.3 to $\sim$0.45 when 50\% of the CO and CO$_2$ budgets were locked into water ice \citep[see Fig.~2 of][]{ligterink2024_trapping}. Their work makes clear the strong influence of volatile entrapment on volatile distributions in protoplanetary discs, although they performed these calculations for a static disc, i.e. without pebble drift and viscous disc evolution.

Understanding how the C/O ratio changes due to volatile entrapment cannot be fully appreciated without pebble drift due to the important role it plays in the chemical evolution of protoplanetary discs \citep[e.g.][]{piso2015_desorption_dist, oberg&bergin2016_C/O_drift, booth2017_giant_enrich}. Dynamical disc models consistently show that the drift of pebbles and the diffusive spreading of vapour can cause significant and time-dependent differences to static disc models when it comes to quantities such as the C/O ratio \citep[e.g.][]{molliere2022_exoplanet_atmo}, a quantity that has previously been used to trace planet formation \citep{molliere2020_trace_planet_CtoO, zhang2021_super_jupiter_atmo, bitsch2022_wasp77Ab, hoch2023_CtoO_formation}. Acquiring a strong grasp on the evolution of the C/O ratio in these discs where planets may be forming is therefore crucial to a comprehensive picture of planet formation.

Recent work by Armitage et al. (in review) found that high-resolution ALMA observations of HD 163296 and MWC 480 could be well-explained by the entrapment of CO in water ice when coupled with radial drift and pebble growth: the observations revealed centrally-peaked CO enhancement profiles, indicating a larger abundance of CO gas in the inner disc ($\lesssim 50$ au) compared to near the CO snowline ($\sim 70$ au).

In this paper, we present a detailed analysis of radial drift models coupled with volatile entrapment in a smooth, viscously evolving protoplanetary disc for the first time to explore how this microphysical phenomenon can have macroscopic effects on the heavy element content available for forming planets. We focus here on CO entrapment inside water ice specifically and how it can significantly impact carbon and oxygen ratios.

This work is structured as follows: we briefly introduce our viscous disc model in section~\ref{subsec:disc model}; we discuss how we implement volatile entrapment in our model in section~\ref{subsec:trapping model}; we present our chemical budget assumptions in section~\ref{subsec:chemical budget}; we present our key results in section~\ref{sec:results}, where we investigate carbon and oxygen ratios (C/O, C/H, O/H), ice compositions; we then discuss the implications of volatile entrapment for the delivery of volatiles to the inner 0.5 au, planets, and observations in section~\ref{sec:discussion}. We also consider the entrapment of other volatiles. Finally, we summarise our work and key conclusions in section~\ref{sec:conclusions}.

\section{Methodology}
\label{sec:methodology}

\subsection{Disc Evolution}
\label{subsec:disc model}
To simulate the time evolution of a protoplanetary disc, we utilise the 1D volatile transport model \texttt{chemcomp}\footnote{\texttt{https://github.com/AaronDavidSchneider/chemcomp}}  \citep{schneider&bitsch2021a_chemcomp}. This code features dust growth and radial drift \citep[based on the \texttt{two-pop-py} code from][]{birnstiel2012_twopoppy} as well as the transport (via gas advection and diffusion), sublimation, and condensation of volatile species. 

The disc is initialised with a gas distribution following the self-similar profile from \citet{lynden-bell&pringle1974}:

\begin{equation}\label{eqn:initial gas profile}
    \Sigma_{\rm{gas}}(r,t=0)= \frac{M_{\rm{disc}}}{2 \pi r_{c}^{2}} \left( \frac{r}{R_{c}} \right)^{-1} \exp \left[- \left(\frac{r}{R_{c}} \right) \right]
\end{equation}

\noindent for disc gas mass $M_{\rm{disc}}$, radius $r$, and characteristic radius $r_c$. The disc then follows a viscous evolution model \citep{lynden-bell&pringle1974, bell1997_viscous_evol} with viscosity described by the parameter $\alpha$ \citep{shakura&sunyaev1973}. We assume that the solid material distribution initially follows equation \ref{eqn:initial gas profile}, but scaled by the dust-to-gas ratio $Z$, which we set to 0.01 in this work:

\begin{equation}
    \Sigma_{\rm{solids}}(r,t=0) = Z \cdot \Sigma_{\rm{gas}}(r,t=0)
\end{equation}

As the disc evolves in time, dust grains grow into pebbles and drift radially inwards, carrying their icy volatiles through snowlines and depositing the volatiles into the gas-phase; this modifies the disc composition and dust-to-gas ratio from the disc's initial state with both radius and time. The full details of the disc evolution and dust growth models can be found in \citet{birnstiel2012_twopoppy} and \citet{schneider&bitsch2021a_chemcomp}.

We model a disc around a solar mass star, with the parameters (e.g. $M_{\rm{disc}}$, $R_{\rm{c}}$ etc.) used in this work detailed in Table~\ref{tab:model parameters}, where we also note the mass ratio of trapped CO ice to water ice. Radiative transfer models by \citet{bergner2024_jwst_ice} assume a trapped-CO-to-water mass ratio of $\sim$15\%, comparable to 7 to 22\% in the models we use here; we discuss this link further in section~\ref{subsec:observations}.

\begin{table}
\centering
    \renewcommand{\arraystretch}{1.2}
    \begin{tabular}{  ccccccc |}
        \hline     
        \begin{tabular}{@{}c@{}}
            $M_{\rm{disc}}$\\
            ($M_{\odot}$)\\
        \end{tabular} & 
        \begin{tabular}{@{}c@{}}
            $R_c$\\
            (au)\\
            \end{tabular} & 
        \begin{tabular}{@{}c@{}}
            $v_{\rm{frag}}$\\
            (cms$^{-1}$)\\
        \end{tabular}
        & $\alpha$ & \COtrap & $\dfrac{M_{\mathcal{T_{\rm{CO}}}}}{M_{\rm{H_2 O}}}$\\
        \hline
        \hline
        0.1 $M_{\odot}$ & 150 & 500 & 10$^{-3}$, 10$^{-4}$ & 0\% & 0\\
        `` & `` & `` & `` & 20\% & 0.072\\
        `` & `` & `` & `` & 40\% & 0.143\\
        `` & `` & `` & `` & 60\% & 0.215\\
        \hline
    \end{tabular}
    \caption{Parameters of the models shown in this work for a solar-type star ($M_{\star}=M_{\odot}$, $L_{\star}=L_{\odot}$), which form the our fiducial model. Each model is run for $\alpha=10^{-4}$ and $10^{-3}$ (`low' and `moderate' viscosity, respectively). \COtrap denotes the trapping fraction of CO in water ice, which corresponds to a mass ratio of the trapped CO ice to water ice, i.e. $M_{\mathcal{T_{\rm{CO}}}}/M_{\rm{H_2 O}}$.}
    \label{tab:model parameters}
\end{table}

The sublimation and condensation of volatiles follows the prescription detailed in \citet{schneider&bitsch2021a_chemcomp}, with the temperature profile dictating the location of the volatile snowlines. Our temperature model follows that of e.g., \citet{armitage2013_astrophysics}, the details of which are outlined in Appendix B of \citet{schneider&bitsch2021a_chemcomp}. This prescription includes non-evolving viscous heating and stellar irradiation, leading to static snowlines throughout the evolution of our models.\footnote{When compared against \texttt{chemcomp} models with evolving viscous heating, the water snowline moves by less than 0.2 au over 5 Myr; we therefore opt for a static temperature profile for computational ease.} Each volatile sublimates at its corresponding snowline, except for volatiles that are trapped in water ice.

We reference several models throughout this work, which feature different volatile entrapment efficencies and viscosity parameters $\alpha$. We refer to the `moderate viscosity' and `low viscosity' model as the models with $\alpha=10^{-3}$ and $10^{-4}$ respectively. For each model, we present a `trapping' and `no-trapping' case, where the trapping models have different trapping efficiencies. We discuss the details of the CO trapping model in section~\ref{subsec:trapping model}.

\subsection{Volatile Trapping Model}
\label{subsec:trapping model}

\begin{figure*}\label{fig:schematic}
	\includegraphics[width=\textwidth]{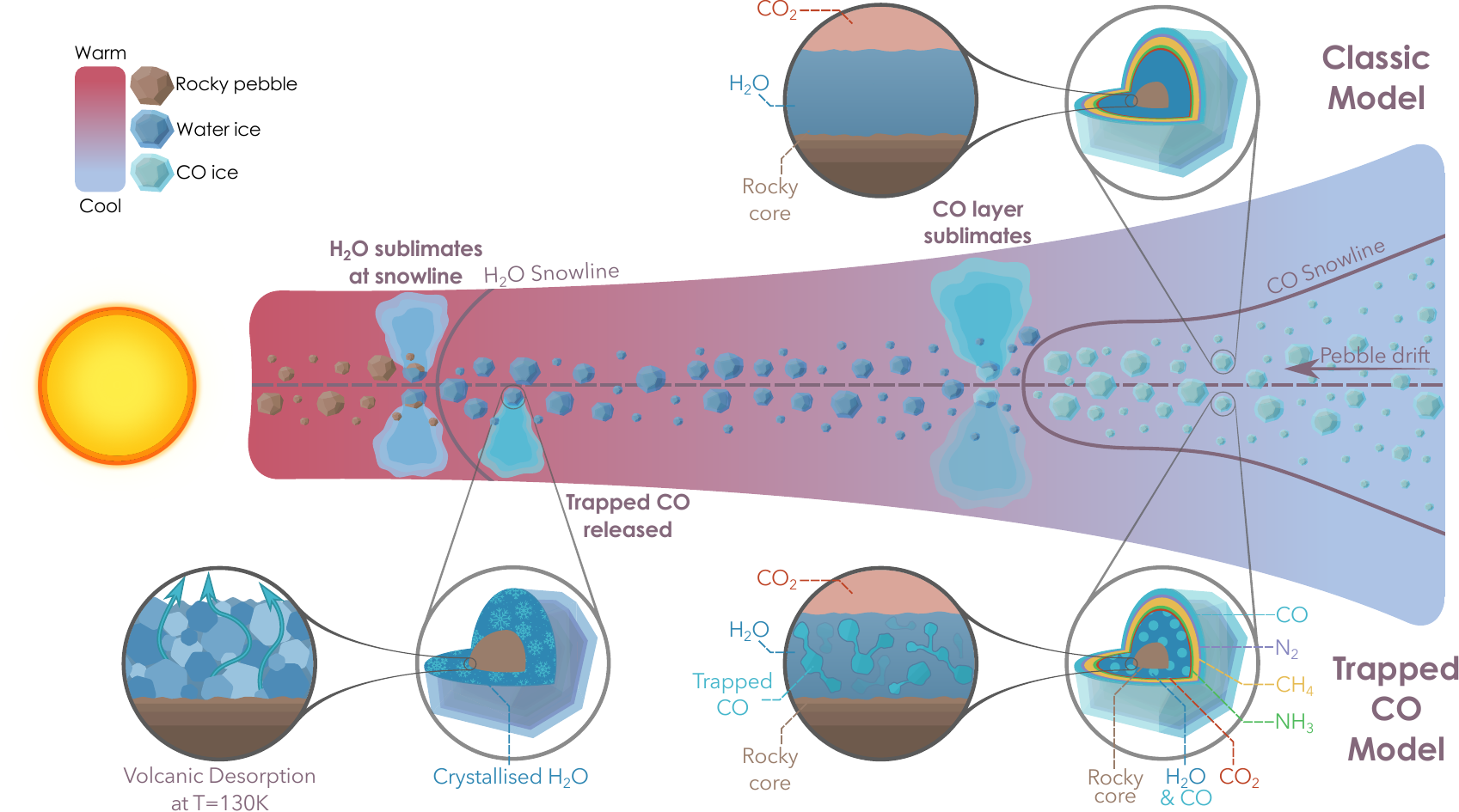}
    \caption{Schematic illustrating a protoplanetary disc  showing a classic ice layering pebble model (top half), and a CO entrapment model (bottom half) where CO is trapped in water ice. Some of the volatile species used in this work are labelled, where the `rocky core' consists of material with a sublimation temperature $T>300\rm{K}$. Pebbles coagulate, grow, and settle towards the midplane as they drift radially inwards. As they pass the CO snowline, the CO layer sublimates; the Classic Model loses all of its CO budget to the gas phase, whereas the Trapped CO Model retains a portion of the CO in water ice; the lower CO abundance in the Trapped CO Model is shown by a more transparent cloud of vapour. As the pebbles drift further, the water ice crystallises at $T=130\rm{K}$ and releases CO contained within the water ice in the trapped CO model via volcanic desorption; nothing occurs in the Classic Model. The water layer then sublimates at the water snowline at $T=150\rm{K}$ in the same way in both models.}
    \label{fig:trapping schematic}
\end{figure*}

We illustrate the key physics of our disc evolution and volatile trapping model in Fig.~\ref{fig:trapping schematic}, showing two scenarios:

\begin{enumerate}
    \item Our no-trapping model, where volatiles are discretely layered and thermally desorb at their snowlines, according to their sublimation temperature.
    \item Our trapping model, where a fraction of the CO reservoir is trapped deeper in the water ice mantle; this is released via `volcanic desorption' at 130K outside the water snowline.
\end{enumerate}

Both of these scenarios feature volatiles modelled as discrete layers of ice atop a rocky core, forming an icy pebble, but our trapping model contains CO ice entrapped in pores inside the amorphous water ice mantle \citep{collings2003_CO_trapping}. In both models, a pure CO layer sublimates as pebbles drift through the CO snowline. When they reach a temperature of 130K, which occurs just outside the water snowline, the amorphous water ice crystallises and releases trapped CO via volcanic desorption \citep{burke&brown2010_ice_desorption}, providing a second location where CO is released. This CO vapour will diffuse both inward and outward.

We define the CO trapping efficiency \COtrap as the fraction of the CO abundance that is locked into water ice at every radius at $t=0$; note that \COtrap is \textit{not} defined by mass. That is, \COtrap=50\% does not mean that 50\% of the CO mass is locked into water ice, but instead 50\% of the CO abundance. This is a consequence of the disc mass being non-uniformly distributed at $t=0$ according to equation~\ref{eqn:initial gas profile}, but the CO/H abundance being constant everywhere. For example, if CO/H $=0.25 \times $ C/H, then \COtrap$=20\%$ corresponds to an abundance of $0.05 \times \rm{C/H}$ being released at the volcano line and $0.2 \times \rm{C/H}$ being released at the CO snowline.

Throughout this work, we refer to the location of volcanic desorption as the `\textit{volcano line}' to distinguish it from snowlines where thermal desorption occurs. We additionally refer to CO entrapped in water ice as `trapped CO' and CO that desorbs at the CO snowline as `pure' or `regular' CO ice.

In the following sections, we outline how we model the trapping of CO inside water ice in \texttt{chemcomp} and justify the assumptions we make throughout this work. 

\subsubsection{Lessons from the Lab - Model Assumptions}
\label{subsubsec:lab lessons}

Laboratory experiments investigating the adsorption of vapour onto substrates have examined a range of parameters associated with trapping in water ice, such as volume mixing ratios \citep[e.g.][]{fayolle2011_co2_trap_in_ice} and ice matrix thickness \citep{pesciotta2024_matrix_thick}, relative trapping efficiencies between volatiles \citep{bar-nun1988_amorphous_water_trap, simon2019_co_trap_in_co2}, low-temperature deposition and competitive entrapment in complex mixtures \citep{zhou_competitive_2024}, and the importance of trapping in ice-phase chemistry \citep{fresneau2014_trapping_chemistry}. Experiments have also investigated the entrapment of different volatiles (e.g. N$_2$, Ar, CO \citep{bar-nun2007_N2_Ar_CO_trap}, CH$_4$ \citep{simon2023_comet_entrapment, zhou_competitive_2024} and H$_2$S \citep{santos2025_h2s_trapping}), but we focus on CO in this work. Here, we detail select results to justify several assumptions throughout this work.

Firstly, we focus on CO entrapment inside water ice as this provides the largest difference between desorption locations (20K versus 130K) compared to other molecules. We discuss entrapment inside CO$_2$ ice and trapping other molecules in section~\ref{subsec:additional volatiles}.

Determining the amount of CO that is trapped inside water ice (or `trapping efficiency', which we denote as \COtrap) is a core component of our model. There has been extensive laboratory study of how much CO can be locked inside water ice \citep[e.g.][]{bar-nun1985_ice_trapping, pesciotta2024_matrix_thick}, which itself depends on the host-to-volatile ratio \citep{fayolle2011_co2_trap_in_ice, pesciotta2024_matrix_thick}. Recent work by \citet{pesciotta2024_matrix_thick} indicates that trapping efficiency is not significantly impacted by the thickness of the water ice layer trapping CO beyond 50 monolayers, and a range of grain sizes are expected to be able to efficiently trap volatiles. We therefore assume that the trapping efficiency is independent of grain size in our work.

Understanding the exact trapping efficiency of volatiles in water ice is a complex task, with different conditions and results present throughout the literature \citep{bar-nun1985_ice_trapping, bar-nun1988_amorphous_water_trap, bar-nun2007_N2_Ar_CO_trap, simon2023_comet_entrapment}. Work by \citet{pesciotta2024_matrix_thick} show that the trapping efficiency of CO depends on the host-to-volatile mixture, with their Fig.~4 demonstrating a possible relationship between the water-to-CO ratio and trapping efficiency. Mixtures with water-to-CO ratios $\gtrsim 2$ show non-negligible levels of entrapment (\COtrap>20\%), reaching \COtrap up to 60\%. For simplicity, we elect to model a range of trapping efficiencies from 0 to 60\% throughout this work; this is in agreement with the trapping efficiencies detailed by \citet{ligterink2024_trapping}.

We additionally assume that trapped CO is formed at low temperatures in the interstellar medium during the collapse of a molecular core from which a protoplanetary disc forms; the low temperatures at which water forms may be sufficient for CO trapping to occur \citep{hama&watanabe2013_asw_review, ciesla2018_noble_trapping, ligterink2024_trapping}. Since water ice is expected to come primarily from the ISM, surviving cloud collapse and therefore retaining mixed-in CO \citep{visser2009_chem_history, oberg2023_chemistry_review}, we assume that the trapping efficiency is constant throughout the disc. During the low-temperature formation of trapped ices, significant competition for binding sites between different volatiles is not expected \citep{zhou_competitive_2024}; therefore, we assume that we do not have to factor in other volatiles when determining \COtrap.

Finally, the release of trapped CO is an irreversible process: not only is it too warm for CO to recondense at the volcano line ($T>20$K), water co-condensing with other volatiles above 90K is ineffective at trapping volatiles \citep{gudipati2023_amorphous_ices}. We therefore prohibit the reformation of trapped CO ice in our model.

\subsubsection{Release of Trapped Volatiles}
\label{subsubsec:modelled volcanic desorption}

Volatile trapping entails the locking of volatiles inside water and CO$_2$ dominated mixtures. Typically, volatiles thermally desorb from the surface of grains at their corresponding desorption temperature \citep{collings2004_thermal_desorb, simon2019_co_trap_in_co2}, but those locked inside water ice remain in the ice phase. These entrapped volatiles are released through either co-desorption with water or `volcanic desorption', where water ice turns from amorphous to crystalline \citep{bar-nun1985_ice_trapping, collings2003_CO_trapping, burke&brown2010_ice_desorption}. Volcanic desorption is expected to occur a few tens of Kelvin below the water desorption temperature \citep{collings2003_CO_trapping, kipfer2024_N2_trapping}, which in our model corresponds to $\sim$0.3 au for the models in Table~\ref{tab:model parameters}.

We elect to model the release of trapped volatiles as a single event at the volcano line rather than incorporating both volcanic desorption and co-desorption. We justify this as follows.

The timescale for gas to diffuse across a distance of $\Delta r$ is given as \citep[e.g.][]{armitage2013_astrophysics}:

\begin{equation}
\label{eqn:viscous timescale}
    t_{\rm{visc}} = \frac{(\Delta r)^2}{\mathcal{D}}
\end{equation}

\noindent for diffusion coefficient $\mathcal{D}\sim\alpha c_{s} H_{\rm{gas}}$ \citep{shakura&sunyaev1973, armitage2013_astrophysics}. Assuming $\alpha=10^{-3}$, a gas mean molecular weight $\mu=2.3$ and a temperature of $\sim$150 K (corresponding to a sound speed of $\sim7\times 10^{5}$cms$^{-1}$), the timescale for viscous diffusion across $\Delta r \sim$0.3 au in the inner disc around a solar mass star is $\sim$25 kyr. This increases to $\sim$250 kyr for $\alpha=10^{-4}$. This means that if trapped volatiles are released at both the volcano line and the water snowline, it will quickly become indeterminable whether the trapped volatiles were released at 130 K or 150 K since these timescales are much shorter than the lifetime of the disc. 

In addition, most of the trapped volatiles are released with volcanic desorption (\citealt{kipfer2024_N2_trapping}, also see Fig.~1 of \citealt{ligterink2024_trapping}) as crystalline water is inefficient in trapping other volatiles \citep{gudipati2023_amorphous_ices}. We therefore model the release of trapped volatiles as a single desorption process occurring at $T=130$ K at the volcano line instead of at the water snowline.

\subsubsection{Implementation in \texttt{chemcomp}}
\label{subsubsec:implementation of trapping}

To simulate volatile trapping in \texttt{chemcomp}, we follow the approach of Armitage et al. (submitted): we introduce a new, separate species into \texttt{chemcomp} that track the CO that is locked inside water ice. We discuss the effect of additional trapping, such as locking CO inside CO$_2$ ice and the entrapment of CO$_2$ inside water ice, in section~\ref{subsec:additional volatiles}.

Since the release of trapped CO is an irreversible process, the CO vapour released at the volcano line can diffuse out to the CO snowline and recondense back onto pebbles as pure CO ice \citep{stevenson&lunine1988_recondense, cuzzi&zahnle2004, ros&johansen2013}. To ensure a proper implementation of this, we convert trapped CO into § upon `sublimation'; this approach is similar to \citet{houge2025_organics}, who investigated the irreversible thermal decomposition of refractory organics. A consequence of this implementation is that the total mass of trapped CO ice monotonically decreases with time.

In the standard \texttt{chemcomp} model, inward drifting icy pebbles release volatiles at their snowlines. Sublimation does not occur instantaneously, but rather occurs over some desorption timescale. The evaporation term in \texttt{chemcomp} is defined such that all icy pebbles sublimate the desorbing volatile species within 10$^{-3}$ au of their snowline \citep[see equation 19 in][]{schneider&bitsch2021a_chemcomp}. However, detailed desorption models show that the desorption rate depends on particle size \citep[e.g.][]{hollenbach2009_desorption_rate, piso2015_desorption_dist, krijt2016_water, booth2017_giant_enrich, stammler2017_CO}, in which case icy pebbles may drift further inside snowlines than $10^{-3}$ au. Indeed, based on Figure 2 of \citet{piso2015_desorption_dist}, a $\sim$1 cm sized particle at a distance of 1 au will desorb over $\sim$0.5 au. We therefore opt for a slightly larger desorption distance of $10^{-2}$ au in this work, which is sufficiently short so as to ensure that all of the grains do not drift into the star as they desorb.

Furthermore, we numerically model the release of trapped CO via volcanic desorption in the same way as the sublimation of e.g. water ice. That is, the sublimation terms for the trapped CO species and the water ice in the \texttt{chemcomp} code follow the same equation \citep[equations 18 and 19 in][]{schneider&bitsch2021a_chemcomp}. The rate of volcanic desorption depends on the rate of water crystallisation, which may depend on the rate at which the water is heated \citep{burke&brown2010_ice_desorption}, although we do not model this explicitly; however, the timescale of water crystallisation is significantly shorter than the age of the disc, so we therefore choose to simulate the release of trapped CO more straightforwardly with the sublimation scheme given in \citet{schneider&bitsch2021a_chemcomp}. This captures the macroscopic effects visible throughout the protoplanetary disc over the course of a few Myr.

\subsection{Chemical Budget}
\label{subsec:chemical budget}

We now detail the volatile and elemental abundances used in our models. Following the approach detailed in \citet{schneider&bitsch2021a_chemcomp}, we establish a partitioning model based on \citet{bitsch&battistini2020_partition}, with modified initial chemical abundances based on \citet{asplund2009_solar_abundances} and molecular volume mixing ratios (shown in Table \ref{tab:full species budget} in Appendix \ref{appendix:chemical partitioning model}). We retain the same chemical abundances as \citet{asplund2009_solar_abundances}, but increase [O/H] from 0 to 0.05 to increase the water abundance in our disc. This elevates the initial water-to-CO ratio to $\sim$4.3 and the O/H abundance to 5.50$\times 10^{-4}$, justifying \COtrap = 60\% (see Sec.~\ref{subsubsec:lab lessons} and Fig.~8 of \citealt{pesciotta2024_matrix_thick}). This super-solar [O/H] still lays within typical stellar abundances \citep{buder2018_GALAH, bitsch&battistini2020_partition} and is therefore a reasonable value. \COtrap$=60\%$ also returns a trapped-CO-to-water ice mass ratio of $\sim$22\%, comparable to radiative transfer models used to match observations \citep[$\sim$15\%,][also see our section~\ref{subsec:observations}]{bergner2024_jwst_ice}.

For completeness when investigating C/O, C/H and O/H ratios, we include all the species given in Table \ref{tab:full species budget} including volatiles and rocks; note that we define `rocks' as species with a sublimation temperature of $T>300\rm{K}$

\section{Results}
\label{sec:results}

Here, we present the results from coupling pebble growth and drift with volatile entrapment. We show the evolution of the mass distribution in our models in section~\ref{subsec:volatile distribution}; how the C/O, C/H and O/H ratios change due to CO entrapment in section~\ref{subsec:elemental ratios}; and finally how the composition of icy pebbles and solid-phase C/O ratio change with radius in  section~\ref{subsec:pebble composition}.

\subsection{Volatile Distribution Evolution}
\label{subsec:volatile distribution}

\begin{figure*}
	\includegraphics[width=\textwidth]{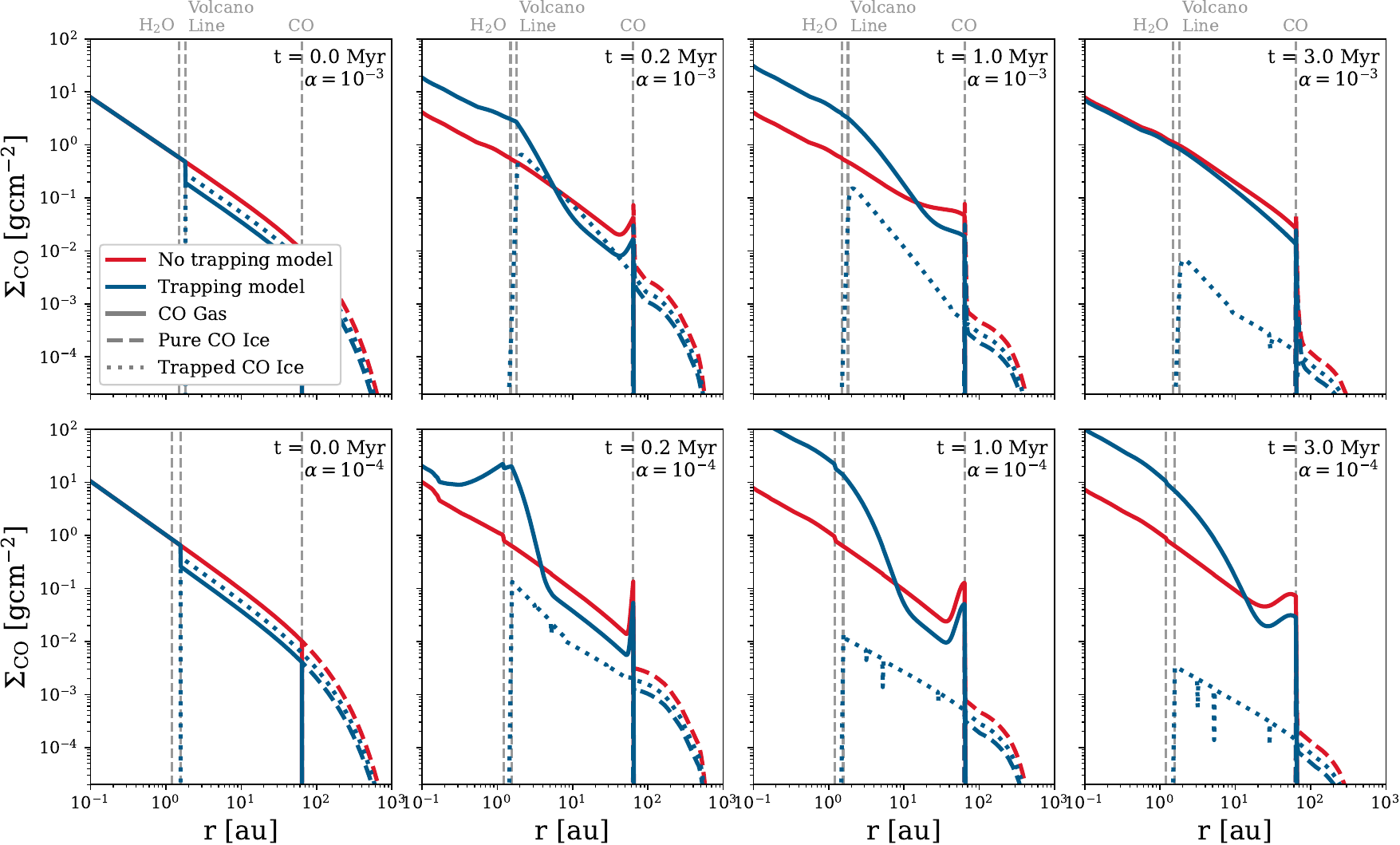}
    \caption{Surface density evolution of pure CO ice (dashed lines) and CO gas (solid lines) for our no-trapping model (blue), and also trapped CO ice (dotted lines) for trapping model (red) of parameters shown in Table~\ref{tab:model parameters} and \COtrap=60\%. Each panel shows a different snapshot in time, at $t=0$, 0.2, 1, and 3 Myr, and the top row shows $\alpha=10^{-3}$ and the bottom $10^{-4}$. CO ice can be reformed from CO condensation at its snowline, whereas trapped CO ice does not reform in our model. Water and CO snowlines are indicated by grey vertical dashed lines and are labelled above the panels in the first row; trapped CO ice is released just before the water snowline. Localised dips near snowlines in the trapped CO ice due to locally altered pebble drift speeds (see text for details). The initial conditions at $t=0$ show reduced CO gas between the water and CO snowlines in the trapping model, as 60\% of the CO abundance has been converted into trapped CO ice. At $t=0.2$ Myr, pebbles deliver CO to the volcano line, which diffuses at different rates depending on $\alpha$. At $t=1$ Myr, trapped CO ice begins to sublimate into regular CO gas and provides a local boost in CO gas. By $t=3$ Myr, the gas has smeared out via diffusion.}
    \label{fig:example evolution}
\end{figure*}

Figure~\ref{fig:example evolution} shows the typical evolution of the no-trapping model against a trapping model with \COtrap=60\% for moderate ($\alpha=10^{-3}$, top row) and low ($10^{-4}$, bottom row) viscosities. The initial conditions at $t=0$ show reduced CO gas surface density outside the volcano line in the trapping model, since a portion of the CO budget has been locked into trapped CO ice. As the disc evolves, inward-drifting icy pebbles first release non-trapped CO at the CO snowline in the outer disc ($\sim$65 au). This CO diffuses both inward and outward, but outward diffusion is halted by re-condensation. 

Nearer to the star, the temperature becomes warm enough for water to crystallise and release trapped CO at the volcano line ($\sim$1.6 au for $\alpha=10^{-4}$ and $\sim$1.8 au for $\alpha=10^{-3}$). Released CO can then diffuse both inward and outward. The release of CO at the volcano line is readily apparent in the $t=0.2$ Myr panel (second column) for $\alpha=10^{-4}$, where a large quantity of CO is released and diffuses radially inward and outward. The same occurs in the $\alpha=10^{-3}$ model, but a lower quantity of CO vapour is released. The CO vapour also diffuses more efficiently due to shorter viscous timescales compared to the $\alpha=10^{-4}$ model. As the vapour diffuses outward to the CO snowline, it is halted by re-condensation, with this outward diffusion shown as the negatively sloped CO gas surface density at $t=0.2$ and 1 Myr. This creates a central peak in the CO abundance, similar to the one inferred for HD 163296 and MWC 480 based on CO isotopologue profiles derived from ALMA observations (Armitage et al. in review). 

Since the turbulence parameter $\alpha$ governs the viscosity, diffusion, and particle size, lower $\alpha$ corresponds to a larger particle size in the fragmentation-limited regime (see equation 8 in \citealt{birnstiel2012_twopoppy}), which the volcano line sits in. This means that lower viscosity discs will have a higher maximum pebble flux near the water snowline \citep{drazkowska2021} and weaker diffusion, resulting in more CO vapour being released at the volcano line, as evident at $t=0.2$ Myr in Fig.~\ref{fig:example evolution}. This leads to a central CO surface density peak, which is larger and longer lived in the low viscosity model \citep{mah2023_vlms_C/O, mah2024_mind_the_gap}. By $t=3$ Myr, this peak has smeared out due to viscous effects in the higher viscosity model.

Finally, we note that there are dips in the trapped CO ice surface density outside the snowlines of other volatiles (CH$_4$, CO$_2$, NH$_3$), most clearly visible at $t=1$ and 3 Myr for the $\alpha=10^{-4}$ model and at $t=3$ Myr for $\alpha=10^{-3}$. This is due to these volatiles diffusing outwards and recondensing onto pebbles, enhancing the local solid-to-gas ratio. This causes pebbles to locally grow and increase their Stokes number, thereby increasing their drift speed. This causes a local decrease in the trapped CO ice surface density.\footnote{This effect also only occurs in the drift-limited-growth regime, which depends on the local solid-to-gas ratio \citep[see equation 18 in][]{birnstiel2012_twopoppy}, whereas fragmentation-limited growth does not. These dips, therefore, only occur in the drift- and not the fragmentation-limited regime, and as the drift limit decreases with time, these peaks begin to appear. Hence, they are mainly present at $t=1$ and 3 Myr, and not 0.2 Myr.}
 
\subsection{Carbon and Oxygen Ratios}
\label{subsec:elemental ratios}

\begin{figure*}
	\includegraphics[width=\textwidth]{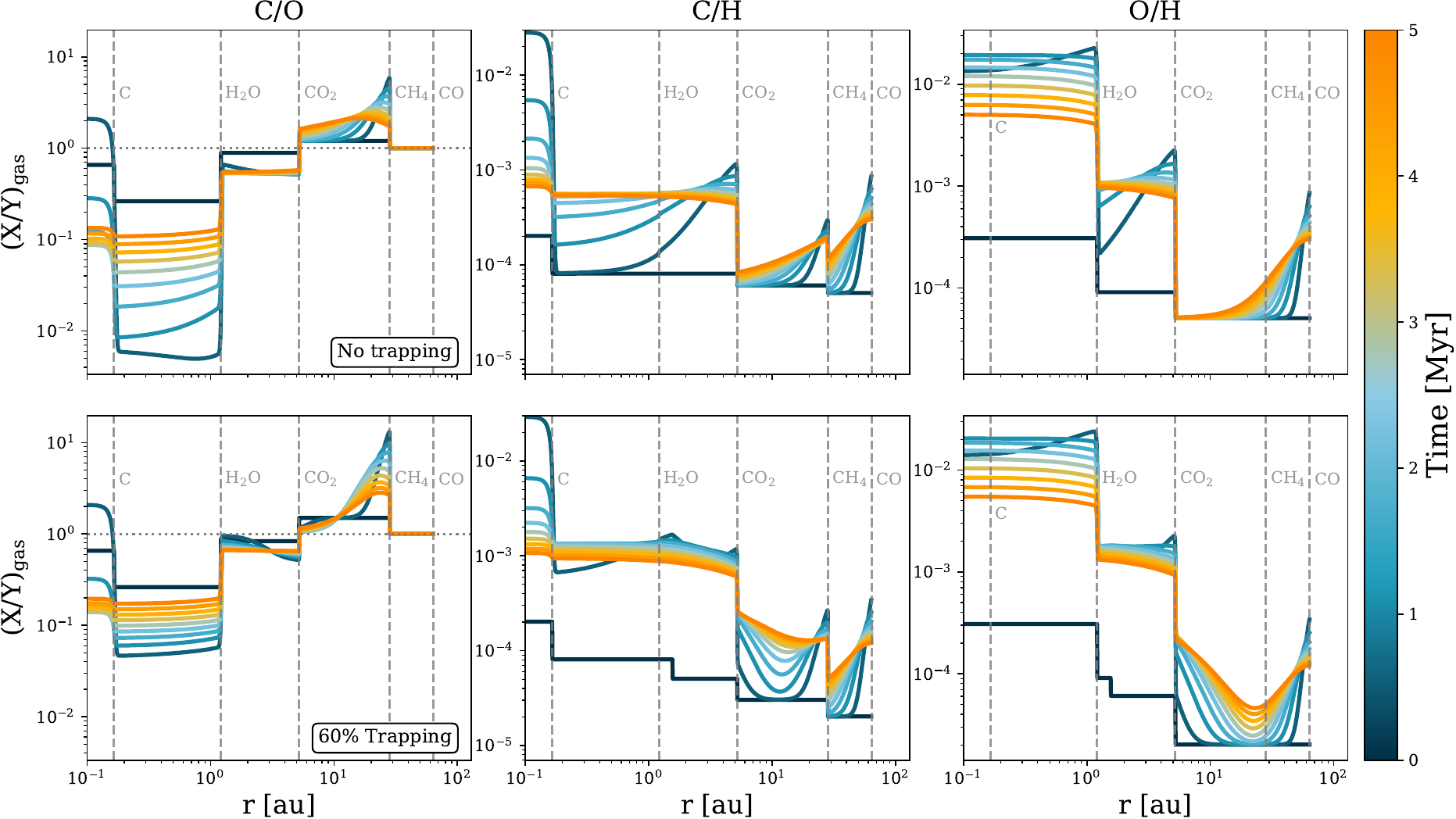}
    \caption{Temporal evolution of the radial gas-phase absolute carbon and oxygen ratios (left: C/O; middle: C/H; right: O/H) for our $\alpha = 10^{-4}$ model, showing both the no-trapping case (top) and \COtrap$=60\%$ trapping case (bottom). C/O=1 is shown as a grey dotted line. Each ratio begins with a static profile, shown by the dark solid line, which then evolves as pebbles grow, drift, and sublimate their ices. The trapping model most significantly alters the C/O ratio in the inner $\sim$3 au and near the CO$_2$ snowline; it also boosts the carbon content inside the CO$_2$ snowline considerably.}
    \label{fig:temporal ratios}
\end{figure*}

A natural extension to the mass distribution of CO in the disc are the carbon and oxygen ratios. Here, we investigate the gas-phase \textit{absolute} C/O, C/H and O/H ratios (i.e. the number ratio) due to all gaseous molecules. Fig.~\ref{fig:temporal ratios} shows the temporal evolution of the radial profiles of C/O (left), C/H (middle), and O/H (right) for our no-trapping and \COtrap~=60\% models for $\alpha=10^{-4}$, showing how the drift and sublimation of pebbles alters the C/O ratio \citep[as also shown by][]{schneider&bitsch2021a_chemcomp}.

CO is released at the its snowline at $\sim$65 au, causing a gas-phase C/O ratio of 1 between the CO and CH$_4$ snowlines. Some of this gas diffuses back across the snowline, recondensing onto pebbles, whilst the remainder advects inward as the gas viscously evolves.  The local sublimation of CO also causes a spike in the C/H and O/H ratios. Introducing other volatiles to the gas-phase causes variations in the carbon and oxygen ratios.

For example, pebbles deliver CH$_4$ to its snowline and significantly boost the C/O and C/H ratios locally as CH$_4$ and CO gas are mixed; the O/H remains unchanged. The sublimation of CO$_2$ and water both reduce the C/O ratio, with CO$_2$ increasing C/H and water boosting O/H. At $\sim$0.2 au, the destruction of carbon grains significantly boosts C/O and C/H.

Introducing entrapped CO redistributes a fraction of the CO (60\% in the model we show in Fig.~\ref{fig:temporal ratios}). Transporting CO to the inner disc via pebble drift is significantly faster than gas advection, meaning entrapment can introduce CO to the inner disc earlier in the disc's lifetime; we discuss this early deliver of CO further in section~\ref{subsec:CO delivery}. As CO is released at the volcano line, it advects inwards and counteracts the reduced C/O from water and CO$_2$ by pushing the C/O to higher values.

Releasing CO in the inner disc also causes a significant boost to C/H, making it almost constant in time inside the CO$_2$ snowline. The C/H and O/H ratios just outside the CO$_2$ snowline also increase, the latter of which is due to the outward diffusion of CO gas from the inner disc.

Our results for the gas-phase C/O ratio contrast to the conclusions of \citet{ligterink2024_trapping}: they argue that the solid-phase C/O ratio changes significantly with different trapping efficiencies (see our section~\ref{subsec:pebble composition}), but the gas-phase C/O ratio is not affected by trapping. Our Fig.~\ref{fig:temporal ratios}, however, shows that trapping has a significant effect on the gas-phase C/O ratio in a dynamical disc; we also find that our gas-phase C/O at $t=0$ (comparable to a static disc) is higher outside the volcano line (see section~\ref{subsubsec:significance parameter} and Fig.~\ref{fig:elemental colourmaps}). We note that we use a slightly different molecular mix from \citet{ligterink2024_trapping}, but our $t=0$ profiles correspond to their results for the same initial mix. Fig.~\ref{fig:temporal ratios} therefore highlights the necessity of dynamical disc models to capture the true nature of volatile evolution due to entrapment.

Exploring the temporal evolution of the radial profile of the absolute gas-phase C/O provides useful insights into the protoplanetary disc's evolution, so we extend this analysis to quantify the impact that CO trapping has on these ratios in the next section.

\subsection{Quantifying the Impact of CO Trapping}
\label{subsubsec:significance parameter}

To evaluate the significance of the impact that CO trapping has on the evolution of these carbon and oxygen ratios, we compare the most extreme trapping efficiency (\COtrap=60\%) to the no-trapping model. This encapsulates the range of C/O, C/H, and O/H values that CO entrapment can cause by examining the most significant effects. We therefore define the significance parameter $\Delta$:

\begin{equation}\label{eqn:significance}
    \Delta(r,t) = \log_{10}\left( \dfrac{(\rm{X/Y})_{\mathcal{T}_{\rm{CO}}=60\%}(r,t)}{(\rm{X/Y})_{\rm{Fiducial}}(r,t)} \right)
\end{equation}

\noindent where absolute gas-phase $\rm{X/Y}$ is the ratio of X (C or O) to Y (O or H), both of which are functions of space and time. $\Delta$ quantifies the significance of CO trapping on the X/Y ratio: when the trapping model provides a larger X/Y, $\Delta$ increases, and reaches unity when the X/Y ratio is a factor of 10 larger due to trapping. Conversely, when the trapping model provides a depletion in X/Y compared to the fiducial model, $\Delta$ decreases and reaches -1 when the X/Y ratio is a factor of 10 smaller due to trapping.

\begin{figure*}
	\includegraphics[width=\textwidth]{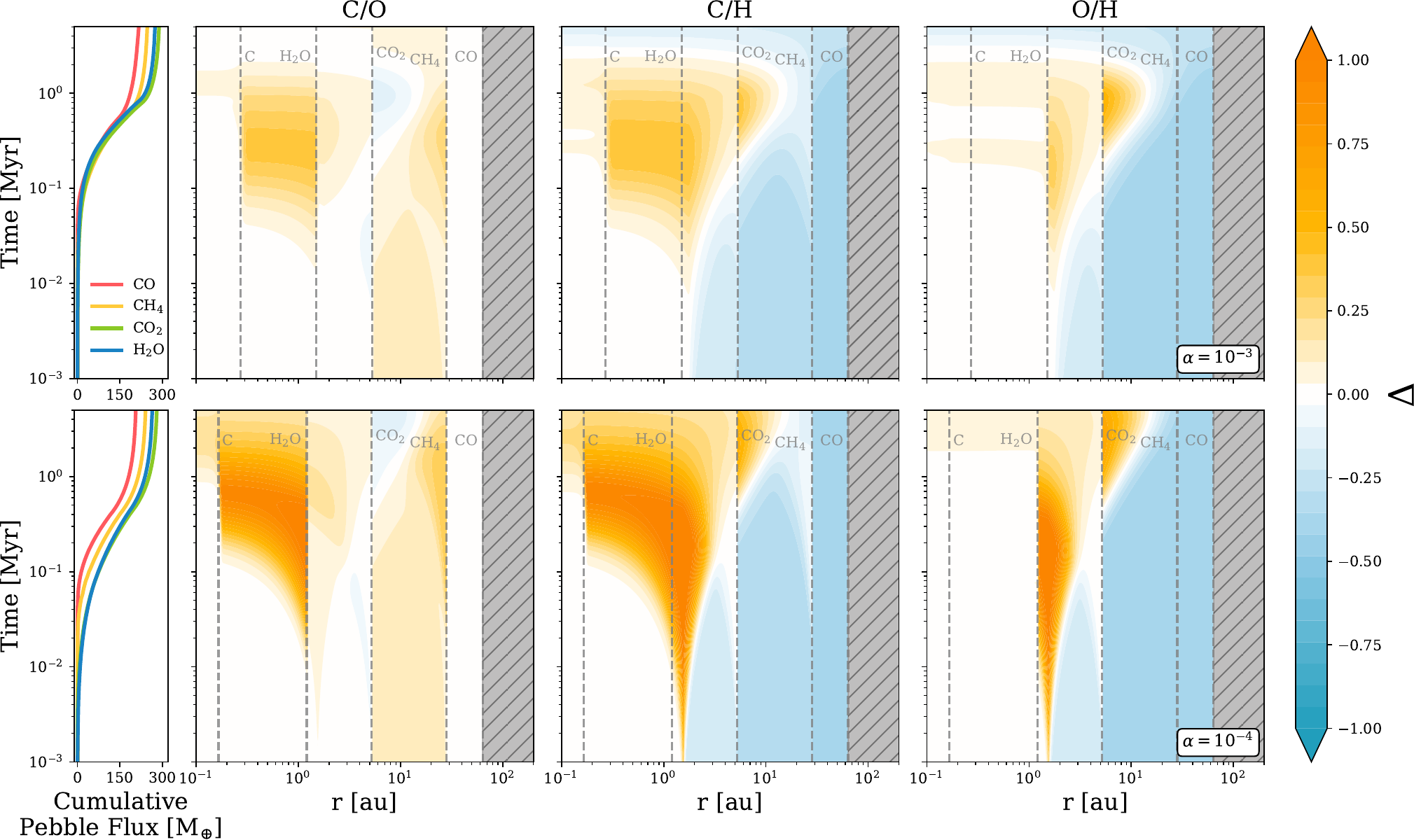}
        \caption{ Significance parameter $\Delta$ (see equation \ref{eqn:significance} in text) quantifying the impact of CO entrapment on the gas-phase absolute C/O (left), C/H (middle) and O/H (right) ratios. The \COtrap=~60\% trapping model is compared against the no-trapping model for $\alpha=10^{-3}$ (top row) and $10^{-4}$ (bottom row); both models include radial drift. Higher on the colourbar (orange) indicates that the trapping model boosts the carbon and oxygen ratio compared to the fiducial, no-trapping case; lower (blue) indicates that trapping cause a reduction.  For example, inside the H$_2$O snowline, the trapping model provides an enhanced C/O at $t\lesssim$ 1 Myr. The grey hatched zone outside the CO snowline indicates that these ratios are undefined due to no gas being present. The cumulative pebble flux is plotted on the left-most panel, with different colours indicating the flux through different snowlines. The biggest enhancement in C/O and C/H around the water snowline coincides with the steepest part of the cumulative pebble flux curves (i.e. when the pebble flux is highest). The volcano line is not labelled for visual clarity, but sits just outside the water snowline.}
    \label{fig:elemental colourmaps}
\end{figure*}

The temporal and spatial evolution of $\Delta$ is shown in Fig.~\ref{fig:elemental colourmaps} for both the $\alpha=10^{-4}$ and $\alpha=10^{-3}$  models, where orange regions indicate a larger X/Y ratio in the \COtrap=60\% trapping model compared to the no-trapping model. We have additionally shown the cumulative pebble flux in an adjacent panel to highlight the relationship between pebble drift and reductions/boosts in the carbon and oxygen ratios.

First of all, we note that every model sees a reduced C/H and O/H outside the CH$_4$ snowline due to the trapped CO being sneaked into the inner disc via icy pebbles; this corresponds with $\Delta\approx -0.3$. CO trapping therefore provides a factor $\sim$2 decrease in the gas-phase C/H and O/H ratios outside the CH$_4$ snowlines.

Fig.~\ref{fig:elemental colourmaps} shows that the trapping model boosts the C/O and C/H ratios inside the water snowline within the first $\sim$1 Myr for both values of $\alpha$. At 0.5 au, $\Delta$ reaches values of $\sim$1 for both C/O and C/H in the low-viscosity model, corresponding to a factor $\sim$10 increase due to CO trapping. In contrast, the moderate viscosity model sees values of $\Delta\approx 0.4$, corresponding a factor $\sim$2.5 increase in C/O and C/H respectively. By 5 Myr, the C/O at 0.5 au returns to the same value as the no-trapping model by $\sim$2 Myr in the moderate viscosity model; the C/H at 0.5 au match the no-trapping model by $\sim$3 Myr and begin to see reduced values in comparison by 5 Myr: the C/H ratio is reduced to $\sim$70\% ($\Delta \approx -0.14$) of the no-trapping model by 5 Myr. The low viscosity model sees the C/O and C/H at 0.5 au reach $\Delta \approx 0.2$, a factor of $\sim$1.6 increase compared to the no-trapping case, by 5 Myr. In all models, the O/H ratio is dominated by water and CO entrapment has negligible effect.

The C/O ratio in the low-viscosity trapping model is significantly higher (factor of $\sim$10) than the fiducial model inside the water snowline due to the additional carbon in CO gas pushing the ratio up; this gas lingers for a few Myr as it slowly dissipates through diffusion and is accreted onto the central star, causing $\Delta$ to slowly decrease. The same effect occurs for the moderate viscosity model, but on shorter timescales.

Introducing a large amount of CO to the inner $\sim$1 au massively boosts the local C/H ratio by introducing significant quantities of carbon. The C/H ratio inside the water snowline increases by a factor of $\sim$10 for the lower viscosity model, and reaching a factor of $\sim$20 ($\Delta \approx1.28$) locally around the volcano line ($\sim$1.5 au). This maximum increase in C/H falls to $\sim$2.5 ($\Delta\approx0.41$) for the moderate viscosity model. This means that CO entrapment causes a significant increase in the carbon abundance inside the water snowline.

The O/H ratio is more resilient against CO trapping in the inner disc, as the O/H ratio is dominated by water. At the volcano line just outside the water snowline, however, the O/H is locally boosted due to the released of CO, although this enhancement is quickly washed away as CO$_2$ gas diffuses and advects inwards in both models and dominates the local O/H ratio. This brings the models in line with each other. Beyond the CO$_2$ snowline, trapping reduces the O/H ratio by a factor of $\sim$2 ($\Delta\approx-0.3$) due to the absence of CO gas.

\subsubsection{Link to Pebble Drift}

The maximum boost in C/O and C/H for both the low- and moderate- viscosity models coincides with the highest pebble flux, i.e. when the rate of change of the cumulative pebble flux is highest. The arrival of drifting pebbles sublimating their water and releasing their trapped CO naturally explains this increase. The low viscosity model features higher pebble fluxes due to faster drift velocities \citep[resulting from the larger fragmentation barrier and pebble sizes][]{birnstiel2012_twopoppy, drazkowska2021}. The boosted C/O and C/H in the first $\sim$Myr of evolution means that young ($\lesssim$ 1 Myr) sources, such as Class 0/I discs or young ones where pebble drift is ongoing, may provide a more obvious footprint of CO trapping inside the water snowline compared to older sources. 

Between the CO$_2$ and CH$_4$ snowlines, the C/O ratio sees an increase of a factor $\sim$2 ($\Delta\approx0.3$) for both values of $\alpha$ due to the reduced amount of CO gas advecting inwards from the CO snowline. There is then a small, local reduction in C/O just outside the CO$_2$ snowline due to the outward diffusion of CO vapour released at the volcano line, occurring at $\sim$1 Myr in the low viscosity model and at $\sim$300 kyr in the moderate viscosity model. The reduction has a value of $\Delta \approx -0.4$ in the low viscosity model and $\sim -0.2$ in the moderate viscosity model, corresponding to factors of $\sim$2.5 and $\sim$1.6 decrease in the C/O due to trapping, respectively. This effect is short-lived in the moderate viscosity model, only surviving for $\sim$ 2 Myr as the CO gas advects inwards, whereas the CO gas survives for longer in the low viscosity model.

The O/H is simultaneously boosted just outside the CO$_2$ snowline for exactly the same reason - the outward diffusion of CO gas releaseed at the volcano line increases the local O/H ratio. These boosted O/H and correspondingly depleted C/O outside the CO$_2$ snowline at later times ($\gtrsim$2 Myr) could therefore be an indicator of CO trapping in older sources (Class II discs, or ones with delayed or halted pebble drift).

Finally, we note that at the C/O and C/H ratios at $\sim$0.2 au are significantly impacted by CO trapping, despite the local sublimation of carbon grains at the `soot line'. This demonstrates that refractory carbon destruction may not be the dominant influence in the gas-phase C/O and C/H; we note, however, that more detailed study of the irreversible sublimation of organic material has significant impacts on the C/O and C/H ratios that we do not detail here \citep{houge2025_organics}. We discuss this further in section~\ref{subsec:irreversible sublimation}.

From Fig.~\ref{fig:elemental colourmaps}, we can see that CO trapping has a much more significant impact in the inner disc and at early times when pebble drift is ongoing ($\lesssim$ 1 Myr). As such, we expect that younger discs - potentially including Class 0/I sources - would have a more noticeable mark from CO trapping in the inner disc, and older discs may bear marks of CO trapping further out from the water snowline. We also discuss the implications of these altered carbon and oxygen ratios for accreting giant and terrestrial planets in section~\ref{subsec:planet formation}.

\subsection{Carbon and Oxygen Ratios Inside the Water Snowline}
\label{subsec:inner disc ratios}

\begin{figure}
	\includegraphics[width=\columnwidth]{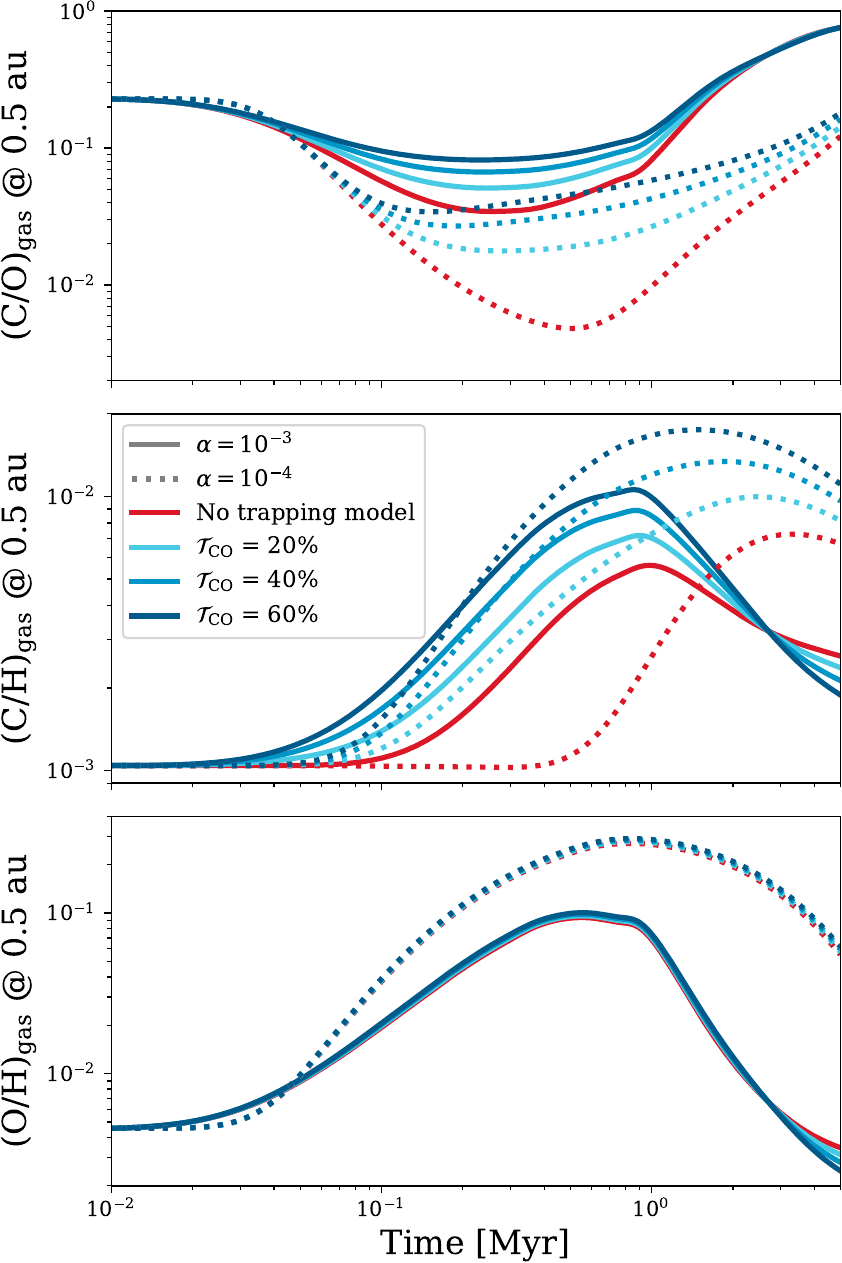}
    \caption{Absolute C/O (top panel), C/H (middle panel) and O/H (bottom panel) ratios in the gas phase at 0.5 au.  The moderate viscosity model (solid lines) quickly sweeps away CO gas released at the volcano line, leading to indistinguishable C/O evolution at $\sim$1 Myr and beyond for each trapping model. Both the high and low viscosity (dotted lines) models show considerable differences in the C/H evolution, although the low viscosity model shows larger differences when CO trapping is introduced. CO entrapment provides little variation in O/H, except for $\sim$ 5 Myr for the moderate viscosity model, when the disc is more depleted in CO for higher trapping efficiencies.}
    \label{fig:inner disc ratios}
\end{figure}

Based on Fig.~\ref{fig:elemental colourmaps}, the C/O and C/H ratios in the inner disc ($\lesssim$1 au) show a considerable difference between the trapping and no-trapping models, warranting a closer look at how these ratios vary with time and trapping efficiency. We therefore explore the temporal evolution of C/O, C/H and O/H at 0.5 au - inside the water snowline but beyond the soot line - for various trapping efficiencies and viscosities in Fig.~\ref{fig:inner disc ratios}.

The C/O ratio of the gas in the top panel of Fig.~\ref{fig:inner disc ratios} shows an increase in C/O for all trapping models compared to the no-trapping model. The C/O in both models begins to decrease at $\sim0.05$ Myr, but the moderate viscosity model shows a smaller decrease. This decrease is tied to the release of water from icy pebbles. The low viscosity model sees a larger decrease in C/O due to the higher pebble flux in low-viscosity discs \citep{drazkowska2021}, delivering more water more quickly, as well as the delayed arrival of carbon-rich gas from the outer disc due to elongated viscous timescales \citep{mah2023_vlms_C/O}. By $\sim$ 1 Myr, the C/O in the high-viscosity discs evolve similarly for all trapping efficiencies as the CO vapour from trapped CO is swept away. In both models, providing CO to the inner disc through volatile trapping is an effective means of increasing the C/O, although the lower viscosity model sees a stronger increase.

In comparison, the C/H ratio of the gas in the middle panel of Fig.~\ref{fig:inner disc ratios} shows similar trends for the low and moderate viscosity cases, although the low-viscosity model viscous timescales are ten times longer. Increasing the trapping efficiency leads to larger C/H ratio with the introduction of additional carbon-bearing gas via pebbles. The low viscosity model shows a more substantial difference between the trapping and no-trapping models due to the larger pebble fluxes and longer viscous timescales, allowing the released CO to linger for longer comapred to the moderate viscosity model. After $\sim$2 Myr, the moderate viscosity model sees an inversion in the model order - that is, at 1 Myr the \COtrap=60\% model has the highest C/H. By 5 Myr, it is the lowest. This is because pebbles can rapidly deliver trapped CO that is then quickly accreted onto the star, compared to CO advecting inwards from the CO snowline to the star; this causes a depletion of carbon in the inner disc. We further discuss the delivery of CO to the inner disc changes with entrapment in section~\ref{subsec:CO delivery}.

Finally, the O/H ratio shown in the bottom panel of Fig.~\ref{fig:inner disc ratios} shows no substantial difference in the O/H evolution over time between trapping models; this is because the inner disc O/H is dominated by water and CO has a very small impact. The only noticeable differences between models can be seen at late times ($\sim$ 5 Myr) for the moderate viscosity model, where the O/H ratio begins to diverge: the no-trapping model shows the largest O/H, likely because more CO gas has been accreted onto the star in the trapping models.

CO entrapment has a noticeable impact on the gas-phase C/O and C/H ratios inside the water snowline, but does not significantly impact the O/H ratio in less than 5 Myr for a disc with $\alpha=10^{-3}$, as O/H is dominated by water. This means that CO entrapment provides a means of introducing more carbon to the inner 0.5 au whilst retaining water. Lower viscosity discs see a more significant impact on their carbon and oxygen ratios.

\subsection{Pebble Composition and Solid-Phase C/O}
\label{subsec:pebble composition}

\begin{figure}
	\includegraphics[width=1\columnwidth]{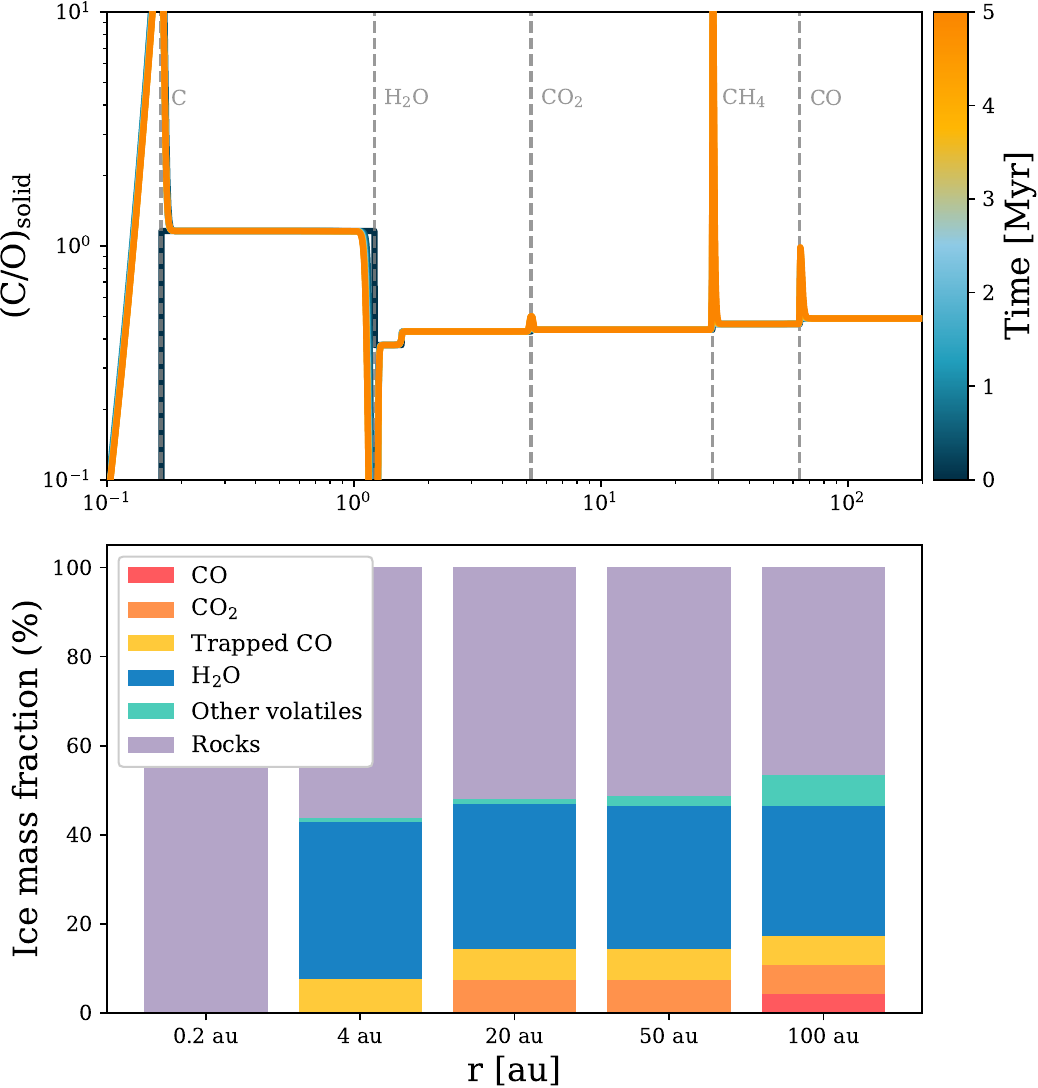}
    \caption{\textit{Top:} temporal evolution of the solid-phase C/O ratio. This ratio changes very little over time, except for local enhancements around snowlines. This is because the mass of volatiles at a given radius all change at the same rate (see text for details). \textit{Bottom:} ice mass fraction of rocks and icy volatiles in pebbles at selected radii at $t=1$ Myr; 100\% mass fraction indicates that the pebbles are made entirely of that species. `Other volatiles' consists of N$_2$, CH$_4$, NH$_3$, and H$_2$S, and `Rocks' includes everything with a sublimation temperature over 300 K. Both plots are for \COtrap=~40\%.}
    \label{fig:pebble composition}
\end{figure}

We now turn to the composition of icy pebbles in the disc. As pebbles drift, their composition changes as they pass through different snowlines and sublimate layers of volatiles. We show the composition of our pebbles at select radii based on the mass fraction of volatiles in Fig.~\ref{fig:pebble composition} for our \COtrap=~60\% model. The pebble is split into its constituent volatiles and `rocks', the latter of we define as having sublimation temperatures over 300 K. We additionally show the temporal evolution of the solid-phase C/O ratio above in the same figure.

We note two key results here. Firstly, the solid-phase C/O ratio is defined almost entirely by the initial conditions of the disc and is effectively static over time, apart from enhancements and depletions local to snowlines; the dark blue curve for $t=0$ is almost identical to the orange curve for $t=5$ Myr. These depletions are due to increased gas densities from recondensation of volatiles. This effectively non-evolution of solid-phase C/O stems from the fact that pebble composition can only be changed by snowlines in our models, since we do not include chemistry \citep[e.g.][]{booth&ilee2019_pebble_chem, eistrup&henning2022_pebble_chem_evol}.

Secondly, we can conclude that the solid-phase C/O ratio between the H$_2$O and CH$_4$ snowlines is consistently higher in the trapping model compared to the no-trapping model due to trapped CO ice, since the solid-phase C/O is dictated by the initial conditions. We show a comparison of this in Fig.~\ref{fig:solid C/O} of Appendix~\ref{appendix:solid C/O}. This is consistent with the findings of \citet{ligterink2024_trapping}.

The ice composition in the bottom half of Fig.~\ref{fig:pebble composition} shows how the ice mass fraction changes with radius. Although we have shown this for $t=1$ Myr, the composition of pebbles will be the same at all times for the reasons discussed above. These radii are chosen to be outside the CO snowline (100 au), and then between subsequent snowlines (50 au is between CH$_4$ and CO; 20 au is between CO$_2$ and CH$_4$; 4 au is between H$_2$O and CO$_2$; and 0.2 au is within the H$_2$O snowline). We package other species in \texttt{chemcomp}, such as N$_2$, CH$_4$, NH$_3$, and H$_2$S (see Table~\ref{tab:full species budget} in Appendix~\ref{appendix:chemical partitioning model}) as `other volatiles'. 

At 4 au, the majority of the icy pebble is rock ($\sim$50\% mass) and water ($\sim$36\%) with $\sim$5\% being trapped CO. By 0.2 au, only rocks remain. At 100 au, outside the CO snowline, CO ice is $\sim$6.2\% of the pebble mass, CO ice trapped in water is $\sim$4.1\%, and water is $\sim$29\% of the pebble mass. As different volatiles sublimate, the mass fraction changes: at 4 au, trapped CO mass fraction has increased to $\sim$5.7\% of the pebble mass and water to $\sim$39\%. By 0.2 au, all volatiles have sublimated, leaving only rocks.

\section{Discussion}
\label{sec:discussion}

\subsection{Two-Phase Delivery of CO to Inside the Water Snowline}
\label{subsec:CO delivery}

Locking CO ice into water ice considerably changes the way that the carbon and CO are transported and distributed in protoplanetary discs. Transporting the CO to the inner disc from the CO snowline can take several Myr via viscous evolution, but pebbles offer a much more rapid means of transporting the CO gas. Here, we investigate the effect of CO entrapment on the abundance of CO in the inner disc (0.5 au) and use it to identify the different delivery mechanisms of CO (icy pebble drift and viscous evolution).

\begin{figure}
	\includegraphics[width=\columnwidth]{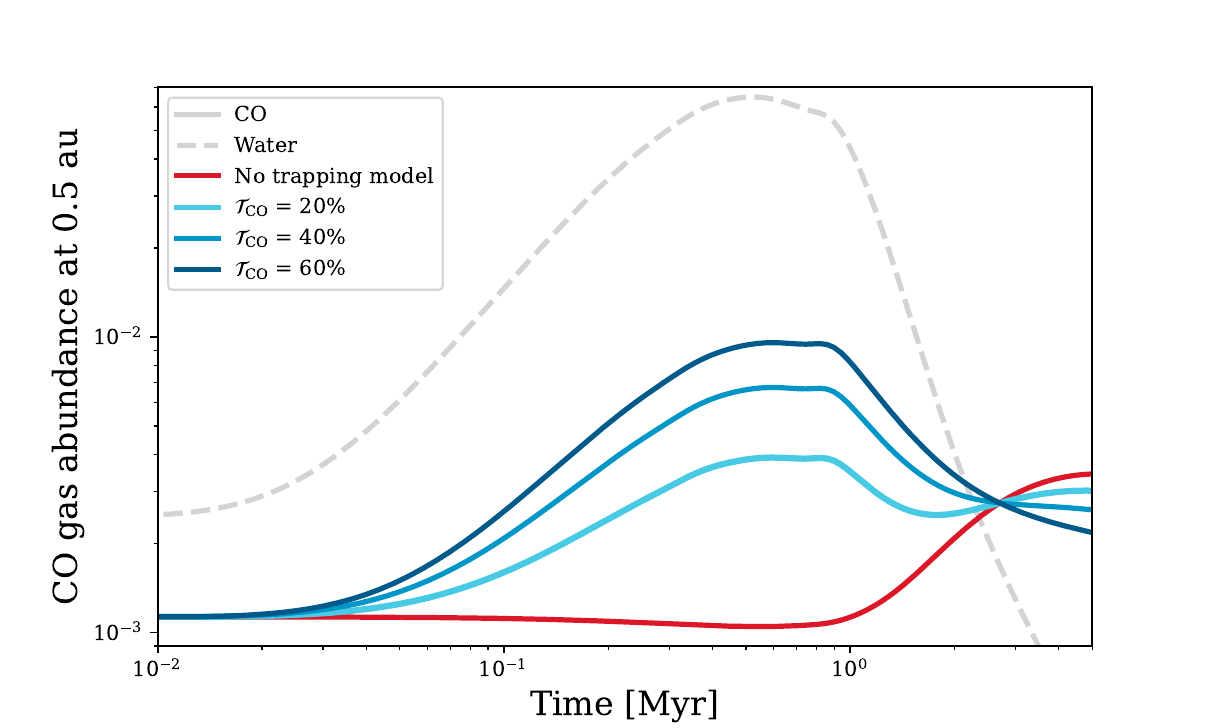}
    \caption{The abundance of CO in the gas-phase (i.e. $\Sigma_{\rm{gas,CO}}/\Sigma_{\rm{gas}}$) at 0.5 au as a function of time for the $\alpha=10^{-3}$ model shown in solid lines; the grey dashed line represents water. Each solid line colour shows the CO abundance for different trapping efficiencies. We show the $\alpha=10^{-3}$ model here as the viscous timescales are sufficiently short to see a two-phase delivery of CO to the inner dics (see text for details).}
    \label{fig:CO abundance}
\end{figure}

The gas-phase CO abundance ($\Sigma_{\rm{gas,CO}}/\Sigma_{\rm{gas}}$) at 0.5 au as a function of time for different trapping efficiencies is shown in Fig.~\ref{fig:CO abundance}; that is, how much of the gas-phase consists of CO vapour. The gas-phase water abundance is also shown in the gray dashed line. We note that the water abundance curve has similar enrichment then depletion phases that have been seen in the literature studying the drift of icy pebbles \citep[e.g.][]{kalyaan2021, kalyaan2023_water_trap, mah2024_mind_the_gap, easterwood2024_water_planet_gaps, sellek2025_co2_drift, houge2025_smuggled_water}.

We show the $\alpha=10^{-3}$ model here and $10^{-4}$ in Appendix~\ref{appendix:CO abundance}; this is because the viscous timescales are short enough in the $\alpha=10^{-3}$ model to identify two phases of CO delivery between the models: the early, icy pebble drift and the late-stage flow of CO via viscous evolution.

The no-trapping model in Fig.~\ref{fig:CO abundance} shows the CO abundance decreasing steadily with time, which coincides with the delivery of water to the inner disc; this causes the relative CO abundance to decrease. After 1 Myr, the CO begins to arrive from the outer disc due to the viscous evolution of the disc. causing the increase in CO abundance \citep{mah2023_vlms_C/O}.

In contrast, the trapping models show a large peak in the CO abundance that coincides with the delivery of water - this demonstrates that the delivery of CO via pebbles is significantly faster than viscous evolution can transport material. Since some CO is trapped in water ice and is released less than 1 au from the water snowline, the CO and water abundance curves follow each other closely. The initial delivery via pebbles causes the CO abundance to peak at $\sim$0.6 Myr.

This peak is then followed by a brief dip and a resurgence in the CO abundance as material flows in from the outer disc in a second delivery phase in the \COtrap$=20\%$ model, where only a small fraction of the CO budget is released at the volcano line. Higher trapping efficiencies dampen the impact of this second-phase delivery, as much of the CO is already present in the inner disc. The net effect of these two mechanisms is that the CO abundance does not change very much beyond $\sim$1 Myr for this particular model.

As the disc evolves beyond $\sim$1 Myr, the water abundance decreases drastically as it is accreted onto the central star, and the CO abundance begins to decrease rapidly before beginning to level off in each model; in the trapping models, this levelling off is due to the outward diffusion of CO aiding its survival in the inner disc by preventing some of the CO from being accreted onto the star.

We note, however, that the outward diffusion of CO depends on the choice of Schmidt number, $\rm{Sc}=\nu/\mathcal{D}$, which we have set to $\rm{Sc}=1/3$ in this work \citep{pavlyuchenkov&dullemond2007_dust_crystallinity}. Understanding the effect of the Schmidt number on the survival of CO is beyond the scope of this work, and refer the interested reader to section~4.4 of \citet{houge2025_organics} for a more detailed investigation into the effect of the Schmidt number on outward transport.

The early stage delivery of entrapped CO creates a distinct imprint on the inner disc CO abundance at $t \lesssim$1 Myr compared to the no-trapping model; in contrast, the late-stage delivery of CO flowing from the outer disc creates a similar CO abundance between the trapping and no-trapping models. Therefore, discs undergoing pebble drift would bear the most obvious hallmarks of volatile entrapment.

\subsection{Planet Formation Implications}
\subsubsection{Planetesimal Compositions}
\label{subsec:planetesimal compositions}

Typically, pebble drift and the sublimation of ices is a key mechanism in determining the chemistry of forming planets \citep{booth2017_giant_enrich, schneider&bitsch2021a_chemcomp, schneider&bitsch2021b_atmospheres}, although the formation of planetesimals can hinder the enrichment that stems from these pebbles \citep{danti2023_giant_atmospheres}. If pebbles are incorporated into planets and planetesimals before they reach the volcano line, the abundance of water vapour (alongside species trapped inside water ice) released in the inner disc ($\lesssim$ 1 au) will be smaller \citep{danti2023_giant_atmospheres}. This effect offers a potential pathway for planetesimals and planets to accrete CO ice as well as depleting the atmosphere of a significant quantity of CO vapour if they form further out from the volcano line.

The post-formation evolution of planetesimals can, however, lead to reduced volatile content down to $\sim$10\% of the initial abundance due to internal planetesimal evolution in less than 1 Myr \citep{lichtenberg&krijt2021_planetesimals}. Trapped CO may then be degassed alongside water ice inside planetesimals, re-releasing the CO into the gas phase. This provides a time constraint: planets accreting planetesimals may have an increased CO content if the planetesimals are accreted within a few hundred kyr of forming, whilst the water ice is still present.

\subsubsection{Implications for Accreting Planet Chemistry}
\label{subsec:planet formation}

The formation of giant planet atmospheres is shaped by the delivery of volatile species via pebble drift and the subsequent volatile distribution \citep{booth2017_giant_enrich, schneider&bitsch2021a_chemcomp, schneider&bitsch2021b_atmospheres, bitsch2022_wasp77Ab, danti2023_giant_atmospheres}, and their atmospheres can provide insights into potential formation pathways \citep[e.g.][]{oberg2011_snowlines, penzlin2024_formation_histories}. Understanding the radial variations of C/O is key for understanding exoplanet atmospheres \citep{marboeuf2008_ices_in_exoplanets, oberg2011_snowlines, johnson2012_planetesimal_comps, bergner2024_jwst_ice}; the ice phase is particularly important due to the importance of pebbles in shaping planetary compositions \citep[e.g.][]{schneider&bitsch2021a_chemcomp, schneider&bitsch2021b_atmospheres}. Hence, disrupting the distribution of volatiles in protoplanetary discs will have significant ramifications for the atmospheres of planets that accrete their atmosphere from the modified gas. 

Since planet formation is expected to start in $\lesssim$ 1 Myr \citep[e.g.][]{johansen&lambrechts2017_review, savvidou&bitsch2023_giant_formation} with growing evidence for the start of planet growth at the Class 0/I stage \citep{drazkowska_pp7_2023}, early-forming giant planets accrete much of their atmospheres during the phase where CO trapping enhances the C/O and C/H in the inner disc. As the planet migrates, it will accrete the carbon-rich vapour, which may then be detectable in exoplanet transmission spectroscopy.

Additionally, trapping CO inside water ice permits it to remain in the ice phase during the collapse of the molecular cloud core \citep{visser2009_chem_history} - the same may be true for when pebbles are accreted onto planets. Typically, pebble accretion luminosities cause sufficient heating so as to sublimate volatiles and recycle the gases back into the protoplanetary disc \citep{johansen2021_solar_system, jiang2023_giant_footprint, wang2023_atmos_recycle}. Entrapment may therefore provide a means to increase the CO content of the accreting planet and boost the atmospheric C/H ratio of an accreting planet that would typically sublimate CO ice, but not water ice. 

However, the exact C/O of the atmosphere of forming giant planets will depend on several additional factors, such as disc structure and chemistry \citep{eistrup2016_volatile_evolution, eistrup2018_evolving_C/O, molliere2022_exoplanet_atmo}, post-formation impacts \citep{sainsbury-martinez&walsh2024_comet_impacts}, tertiary planets blocking volatile delivery \citep{eberlein2024_multi_planet_atmospheres}, and the exact migration pathway of the planet. Tracing the exact formation pathway of planets based on their atmospheres is therefore difficult \citep[see also][and references therein]{penzlin2024_formation_histories}. Volatile trapping further complicates tracing planet formation history, and so studying the footprint of volatile trapping on giant planet atmospheres will be the subject of future work.

It is clear that due to the influence of CO trapping on the radial and temporal variation carbon and oxygen ratios, volatile trapping will have a significant impact on the formation of planets. The giant planets passing through the orange affected regions in Fig.~\ref{fig:elemental colourmaps} will have their atmospheres impacted, and the solid core content will be changed when accreting pebbles and planetesimals which have retained their water ice and the volatiles trapped within; the influence of trapping will depend on how and when forming planets begin accretion. We investigate the material that is available for giant planet formation more closely in the next section.

\subsubsection{Heavy Elements and the Formation of Jupiter}
\label{subsubsec:giant planet atmospheres}

\begin{figure}
	\includegraphics[width=1\columnwidth]{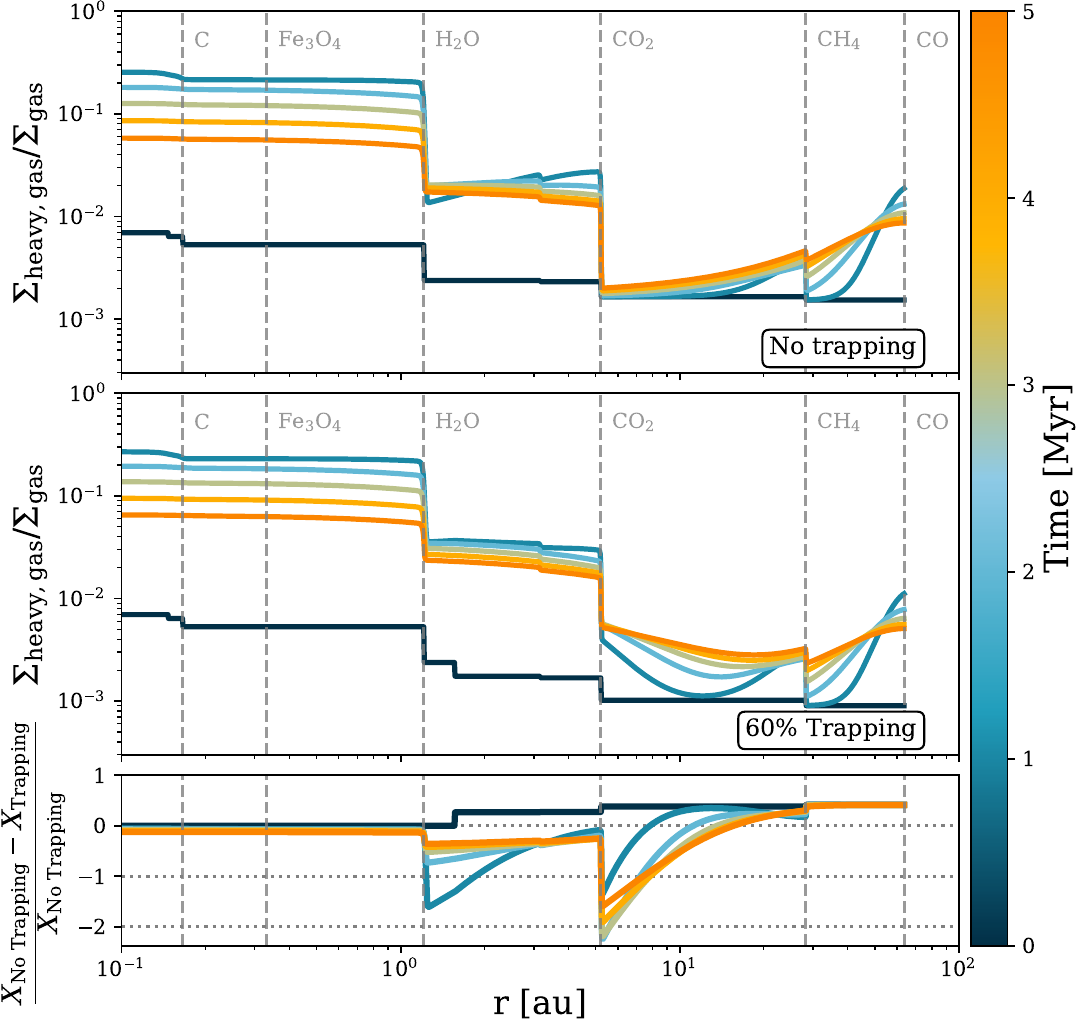}
    \caption{Temporal evolution of the gas-phase heavy element content $X$ for our $\alpha=10^{-4}$ model, showing the no-trapping model (top) and \COtrap$=60\%$ model (middle). The bottom panel shows the fractional difference between the no-trapping model and the trapping model, where a more negative value indicates that the trapping model boosts the heavy element content; a value of -1 corresponds to a boost of 100\% by mass due to CO entrapment. CO trapping provides the biggest fractional change in the heavy element content outside the water snowline and CO$_2$ snowline. The colourbar indicates the evolution of the disc, with blue lines showing $\sim$ 1 Myr and orange 5 Myr.}
    \label{fig:heavy elements}
\end{figure}

The carbon and oxygen ratios of giant planets will be directly influenced by the availability of gas-phase volatiles, which is strongly influenced by the drift of pebbles \citep{piso2015_desorption_dist, booth2017_giant_enrich, schneider&bitsch2021a_chemcomp, schneider&bitsch2021b_atmospheres, bitsch2022_wasp77Ab, bitsch&mah2023_giant_enrichment}. Here, we examine the heavy element content, or volatile abundance with respect to the background gas, of the disc over time. This heavy element content is an indicator of the material available for giant planets to accrete material from the gas. We show the temporal evolution of the gas-phase heavy element content in Fig.~\ref{fig:heavy elements} for our low viscosity disc, and show the case for our moderate viscosity disc in Fig.~\ref{fig:heavy elements appendix} in Appendix \ref{appendix:heavy element content}.  

Planets accreting material interior to the water snowline will accrue all of the volatiles available in the gas phase, and CO trapping makes very little difference here. The bottom panel of Fig.~\ref{fig:heavy elements} shows that inside 1 au, CO trapping provides a decrease in the heavy element content by at most $\sim$10\%; this is achieved at 5 Myr, long after giant planets are likely to have accreted their atmospheres.

The most significant differences in the heavy element content occurs outside the CO$_2$ snowline, where CO trapping depletes the gas-phase CO after 1 Myr for $\alpha=10^{-4}$. At the volcano line, the trapping model provides a $\sim$ 160\% boost in mass of the heavy element content with respect to the no-trapping model at 1 Myr, which decreases quickly to $\sim$75\% by 2 Myr; this is shown by the trough outside 1 au in the bottom panel of Fig.~\ref{fig:heavy elements}, indicating a higher heavy element content in the trapping model. The $\alpha=10^{-3}$ model does not see especially strong enhancements in the heavy element content apart from outside the CO$_2$ snowline inside the first Myr of evolution, where the content is boosted by $\sim100\%$ compared to the no-trapping model. As the disc evolves beyond $\sim$3 Myr, the heavy element content is marginally reduced compared to the no-trapping model.

Work by \citet{schneider&bitsch2021b_atmospheres} found that the drift and sublimation of pebbles was sufficient to explain the Nitrogen content of Jupiter in a non-migrating scenario, although the C/H ratio was found to be too low. Figures~\ref{fig:elemental colourmaps} and \ref{fig:heavy elements} both show that the C/H ratio and heavy element content around the water snowline is significantly boosted due to CO trapping, which may be sufficient to increase the carbon content of Jupiter in their simplified model. \citet{barnett&ciesla2022_jovian_process} suggest that trapping N$_2$ inside more refractory ices may also help explain this enrichment. Incorporating CO trapping into models of Jupiter's formation could therefore provide a better match to observations without needing to invoke migration models \citep[e.g.][]{bitsch2015_pebble_accretion, mousis2019_enriching_jupiter}. However, other formation scenarios suggest that Jupiter formed closer to the N$_2$ snowline \citep{oberg&wordsworth2019_jupiter, bosman2019_jupiter_form}. Fully understanding the role of volatile entrapment in the formation of Jupiter will require further study.

\subsection{Irreversible Sublimation of Carbon}
\label{subsec:irreversible sublimation}

Our results in Fig.~\ref{fig:elemental colourmaps} suggest that CO trapping plays a dominant role in dictating the C/O and C/H ratios inside the water snowline and can increase these ratios significantly. However, our model involves the sublimation and, critically, the `condensation' of carbon grains at the carbon-grain evaporation front at $\sim$0.2 au, whereas this may not be realistic. Recent work by \citet{houge2025_organics} investigates this evaporation front specifically: they replace the carbon grains for organic material, which are irreversibly destroyed at the `soot line' and become some form of hydrocarbon, such as C$_2$H$_2$. This C$_2$H$_2$ then diffuses outwards to its own snowline at $\sim$7 au, inside the CO$_2$ snowline. The net effect of this is to increase the C/O and C/H ratios significantly outside the soot line, which may help explain hydrocarbon-rich inner discs around very low mass stars \citep{arabhavi2025_minds_VLMS}; this effect could dominate over any CO trapping occurring in our models, meaning that any CO released at the volcano line will cause a much smaller fractional increase in the C/O and C/H ratio. A more comprehensive view of the distribution of carbon in the inner disc factoring in different phenomena is required and will be the subject of a future study.

\subsection{Trapping Additional Volatiles: CH$_4$, CO$_2$, N$_2$, Ar}
\label{subsec:additional volatiles}

CO may not be the only volatile to be trapped inside water ice \citep[e.g.][]{bar-nun2007_N2_Ar_CO_trap, simon2023_comet_entrapment, zhou_competitive_2024}, and CO$_2$ may act as another host for volatile trapping \citep{simon2019_co_trap_in_co2, pesciotta2024_matrix_thick}. Including additional volatile trapping, such as CH$_4$ inside water ice, could significantly change the carbon and oxygen ratios across the disc, depending on the trapping efficiency. Similarly, H$_2$S is likely to be very efficiently trapped inside water ice \citep{santos2025_h2s_trapping} and may also influence the C/H and O/H ratios. Including N$_2$ and Ar trapping could provide a way to explain the Nitrogen enrichment of Jupiter in the solar system \citep{ monga&desch2015_jupiter_enrich, mousis2019_enriching_jupiter}, although other explanations such as the dissolution of Jupiter's core of accumulated planetesimals (e.g. \citet{vazan2018_jupiter_evolution, debras&chabrier2019_jupiter_model}; see \citet{helled2022_jupiter_review} for a review), sublimation of drifting pebbles in a non-migrating planet scenario \citep{schneider&bitsch2021b_atmospheres}, forming Jupiter outside the N$_2$ snowline \citep{oberg&wordsworth2019_jupiter, bosman2019_jupiter_form}, and ammonium salts \citep{nakazawa&okuzumi2024_jupiter_ammonia_salts} offer alternative solutions. 

Self-consistently including the trapping of other volatiles in water ice is crucial to understanding the full impact of volatile entrapment in the planet-forming environment and will be the subject of a future study.

\subsubsection{Trapping Inside CO$_2$ ice}

For example, including CO trapping inside CO$_2$ ice and releasing it through co-desorption will reduce the amount that is released at the volcano line. Based on work by \citet{bergner2024_jwst_ice}, the majority of CO trapping in the upper layers of protoplanetary may actually occur inside CO$_2$, leaving a small amount of CO to be trapped inside H$_2$O (see their Fig.~6). Indeed, laboratory experiments by \citet{kipfer2024_N2_trapping} find that more CO is released with CO$_2$ than with water, indicating that the trapping efficiency inside CO$_2$ is higher than in water. Introducing trapping inside CO$_2$ would therefore impact the carbon and oxygen ratios by further redistributing CO closer to the CO$_2$ snowline, where viscous timescales are longer compared to the volcano line. This could mean that the CO survives longer, but this will likely lead to less CO diffusing outward from the volcano line, where viscous timescales are shorter. Understanding the balance between diffusion and viscous accretion will require further study of the combination of CO entrapment in water and CO$_2$ ice.

\subsubsection{Extremely Efficient Trapping with Methyl Mercaptan}

Work by \citet{narayanan2025_methyl_mercaptan} suggests that methyl mercaptan (CH$_3$SH or MeSH) can become entrapped in water ice similarly to CO. When MeSH becomes entrapped alongside other volatiles, such as CO$_2$, MeSH is capable of completely preventing the thermal desorption of CO$_2$ (see their section 4.2). Instead, all of the MeSH and CO$_2$ volcanically desorbs during water crystallisation. If MeSH is present in interstellar ices during the formation of entrapped volatiles, then the distribution of volatiles and their sublimation from icy pebbles will significantly change. MeSH has indeed been found in comet 67P/Churyumov-Gerasimenko \citep{calmonte2016_67P_sulfur} and in the gas-phase across different environments \citep[e.g.][see also references within \citealt{narayanan2025_methyl_mercaptan}]{cernicharo2012_methoxy_gas}, although its abundance around T-Tauri stars remains unknown \citep{narayanan2025_methyl_mercaptan}.

\subsection{Observing Ice Composition in Edge-On Discs}
\label{subsec:observations}

The composition of ices can be inferred from ice absorption spectra as observed by JWST through the use of radiative transfer models \citep{sturm2023a_HH48NE_geometry, sturm2023b_HH48NE_ices, sturm2023c_ice_age, bergner2024_jwst_ice, ballering2025_water_ice_jwst}. Adjusting the composition of ices and the level of mixing of ices produces different absorption spectra shapes in radiative transfer models (e.g. \texttt{RADMC-3D} \citealt{dullemond&dominik2012_RADMC3D}; see section~3.1 of \citealt{bergner2024_jwst_ice} for detailed discussion on the impact of ice mixtures on observed spectra).

The trapped-CO-to-water mass ratio assumed in our models are comparable to those assumed by \citet{bergner2024_jwst_ice} in their radiative transfer models (our 7 to 22\% compared against their $\sim$15\%). Our models also compute the radial ice composition, which depends on disc properties and atomic and molecular abundances (Fig.~\ref{fig:pebble composition}). Since our trapped-CO-to-water mass fraction is similar to observational models, the composition output from \texttt{chemcomp} could act as inputs for radiative transfer models, providing a promising link between the theoretical results of pebble drift models and ice absorption spectra as observed by JWST. Comparing the radiative transfer outputs for trapping and non-trapping scenarios, when coupled with pebble drift outputs, would provide a powerful connection between the optically thick midplane and optically thin surface layers, where the mixed ices are observed by JWST.  

Finally, including additional volatiles (e.g. CH$_4$) in radiative transfer models would provide a negligible impact in the absorption spectra due to low abundances around 5\% \citep{bergner2024_jwst_ice}. Including trapped CH$_4$ in pebble drift models would therefore return results that are likely indistinguishable from other scenarios based on ice absorption spectra.

\section{Conclusions}
\label{sec:conclusions}

We combine the entrapment of CO inside water ice with the 1D viscous evolution code for protoplanetary discs \texttt{chemcomp} \citep{schneider&bitsch2021a_chemcomp} to couple volatile entrapment with pebble growth and drift for the first time. To do so, we introduce a new species into \texttt{chemcomp} with a sublimation temperature of 130K, representing the non-thermal volcanic desorption of entrapped CO from crystalline water ice at what we call the `volcano line'. Our models show that entrapment has a non-negligible effect on the distribution of carbon and oxygen in dynamical discs and should therefore be considered by future volatile distribution studies.

We explore the evolution of the distribution of CO in a low- ($\alpha=10^{-4}$) and moderate-viscosity ($\alpha=10^{-3}$) disc and investigate how this distribution impacts the gas-phase C/O, C/H, and O/H ratios over time, as well as in the inner disc (0.5 au). We also inspect the solid-phase C/O ratio and ice-phase composition of pebbles, and discuss how CO trapping could assist with CO depletion in protoplanetary discs. We further discuss the implications for the chemistry of accreting giant and terrestrial planets, and draw links between our work and observational models from \citet{bergner2024_jwst_ice}. Our main conclusions are as follows:

\begin{itemize}
    \item Pebble drift and the entrapment of CO inside water ice significantly impacts the distribution of carbon and oxygen in dynamically evolving discs, altering their composition.
    \item CO entrapment has the most obvious fractional impact on the C/O and C/H ratios inside the water snowline and at early times that coincide with pebble drift ($t\lesssim1$ Myr). Therefore, the most obvious footprints of volatile entrapment could be expected to occur in young sources.
    \item The C/O ratio is most significantly impacted by CO entrapment outside the water and CO$_2$ snowlines. O/H ratios are affected primarily outside the CO$_2$ snowline, being boosted by a factor of a few, but is otherwise resilient against CO entrapment.
    \item A significant amount of carbon is introduced to inside of 10 au compared to models without trapping at $t<1$ Myr. In the inner 0.5 au, the C/H ratio is boosted by up to factor of 10 for $\alpha=10^{-4}$ and a factor of $\sim$4.7 for $\alpha=10^{-3}$. This can create carbon-rich inner discs around younger sources whilst retaining a large quantity of water.
    \item Moderate viscosity discs ($\alpha=10^{-3}$) show a considerable difference from lower viscosity ($\alpha=10^{-4}$) ones. Lower viscosity discs see a much larger fractional increase in C/O and C/H due to faster pebble drift and longer viscous timescales allowing CO released at the volcano line to pile up.
    \item CO entrapment causes a two-phase delivery of CO to the inner 0.5 au, firstly due to pebble drift followed by viscous advection of vapour from the CO snowline. This causes the CO abundance to vary much earlier than in classical pebble drift models.
    \item Forming planets and planetesimals will have access to otherwise-unavailable solid-phase CO ice, boosting their carbon content. Similarly, the composition of atmospheres of giant planets accreting their atmospheres from this carbon-rich gas will differ to scenarios without trapping if they accrete their atmospheres inside the CO$_2$ snowline.
    \item Introducing the entrapment of further molecules, such as CO$_2$, N$_2$, and CH$_4$ may further impact carbon and oxygen ratios and should be considered in a future study.
\end{itemize}       

Our work demonstrates that the combination of pebble growth, drift, and volatile entrapment can significantly impact disc composition, with the most notable effects occurring early in the disc lifetime and shortly after the end of pebble drift. This will influence the composition of the planets that form in these discs. Future studies on the distribution of volatiles and pebble drift should consider volatile entrapment due to the significant impact it has on the evolution of disc composition.

\section*{Acknowledgements}

The authors would like to thank the anonymous reviewer for their helpful comments and suggestions that helped improve the manuscript. JW is funded by the UK Science and Technology Facilities Council (STFC), grant code ST/Y509383/1. A.H. acknowledges funding from the Carlsberg Foundation (Semper Ardens: Advance grant FIRSTATMO, PI: Anders Johansen). This publication is based upon work from COST Action CA22133, supported by COST (European Cooperation in Science and Technology).\footnote{www.cost.eu} The authors would like to thank Giulia Perotti, Karin Öberg, Pooneh Nazari, Till Kaeufer, Suchitra Narayanan, and Edwin Bergin for fruitful discussions and insightful conversations. This work has made use of the Python packages \texttt{numpy} \citep{harris2020} and matplotlib \citep{hunter2007}.

\section*{Data Availability}

The data underlying this article will be shared on reasonable request to the corresponding author.

\bibliographystyle{mnras}
\bibliography{bibfile}

\appendix

\section{Chemical Partitioning Model}
\label{appendix:chemical partitioning model}

\begin{table}
\centering
    \label{tab:elemental budget}
    \renewcommand{\arraystretch}{1.2}
    \begin{tabular}{ | cc |}
        \hline     
        Element & Abundance \\
        \hline
        \hline
        He/H & 0.085 \\
        O/H & 5.50$\times10^{-4}$ \\
        C/H & 2.69$\times10^{-4}$ \\
        N/H & 6.76$\times10^{-5}$ \\
        Mg/H & 3.98$\times10^{-5}$ \\
        Si/H & 3.24$\times10^{-5}$ \\
        Fe/H & 3.16$\times10^{-5}$ \\
        S/H & 1.32$\times10^{-5}$ \\
        Al/H & 2.82$\times10^{-6}$ \\
        Na/H & 1.74$\times10^{-6}$ \\
        K/H & 1.07$\times10^{-7}$ \\
        Ti/H & 8.91$\times10^{-8}$\\
        V/H & 8.59$\times10^{-9}$ \\ 
        \hline
    \end{tabular}
    \caption{List of elements and abundances from \citet{asplund2009_solar_abundances} used in our model. We modify O/H to be $10^{0.05}$ times larger.}
\end{table}

\begin{table*}
    \label{tab:full species budget}
    \renewcommand{\arraystretch}{1.2}
    \begin{tabular}{ | ccc |}
        \hline     
        Species & T$_{\rm{cond}}$ (K) & $v_{\rm{Y}}$ \\
        \hline
        \hline
        CO & 20 & 0.25 $\times (1-$\COtrap$) \times$ C/H \\
        N$_2$ & 20 & 0.45 $\times$ N/H \\
        CH$_{4}$ & 30 & 0.05 $\times$ C/H \\
        CO$_{2}$ & 70 & 0.1 $\times$ C/H \\
        NH$_3$ & 90 & 0.1 $\times$ N/H \\
        Trapped CO & 130 & 0.25 $\times $\COtrap$ \times$ C/H \\
        H$_{2}$S & 150 & 0.1 $\times$ S/H \\
        H$_{2}$O & 150 &
            \begin{tabular}{@{}c@{}}O/H - (3 $\times$ MgSiO$_3$ + 4 $\times$ Mg$_2$SiO$_4$/H + CO/H + 2$\times$CO$_2$/H + 3$\times$Fe$_2$O$_3$/H + \\
            4$\times$ Fe$_3$O$_4$/H + VO/H + TiO/H + 3$\times$Al$_2$O$_3$ + 8$\times$ NaAlSi$_3$O$_8$ + 8$\times$KAlSi$_3$O$_8$ \end{tabular} \\
        Carbon grains & 631 & 0.6 $\times$ C/H \\
        FeS & 704 & 0.9 $\times$ S/H\\
        Mg$_2$SiO$_4$ & 1354 & Mg/H \\
        Fe$_2$O$_3$ & 1357 & 0.5 $\times$ (FeH - 0.9 $\times$ S/H) \\
        VO & 1423 & V/H \\
        MgSiO$_3$ & 1500 & Mg/H - 2 $\times$ (Mg/H - (Si/H - 3 $\times$ K/H - 3 $\times$ Na/H)) \\
        Al$_2$O$_3$ & 1653 & 0.5 $\times$ (Al/H - K/H + Na/H)) \\
        TiO & 2000 & Ti/H \\
        \hline
    \end{tabular}
    \caption{Full species (volatiles and rocks) budget for our fiducial and all trapping models, based on Table 2 of \citet{schneider&bitsch2021a_chemcomp}, showing the mixing ratios $v_{\rm{Y}}$ of particular species for a particular model. Carbon grain abundance based on \citet{gail&trieloff2017_carbon}. Condensation temperatures are based on \citet{lodders2003_cond_temps}. \COtrap denotes the trapping fraction of CO, which we vary from 0 to 0.6 in steps of 0.2.}
\end{table*}

\section{Solid-phase C/O Comparison for trapping and no-trapping}
\label{appendix:solid C/O}

\begin{figure}
	\includegraphics[width=\columnwidth]{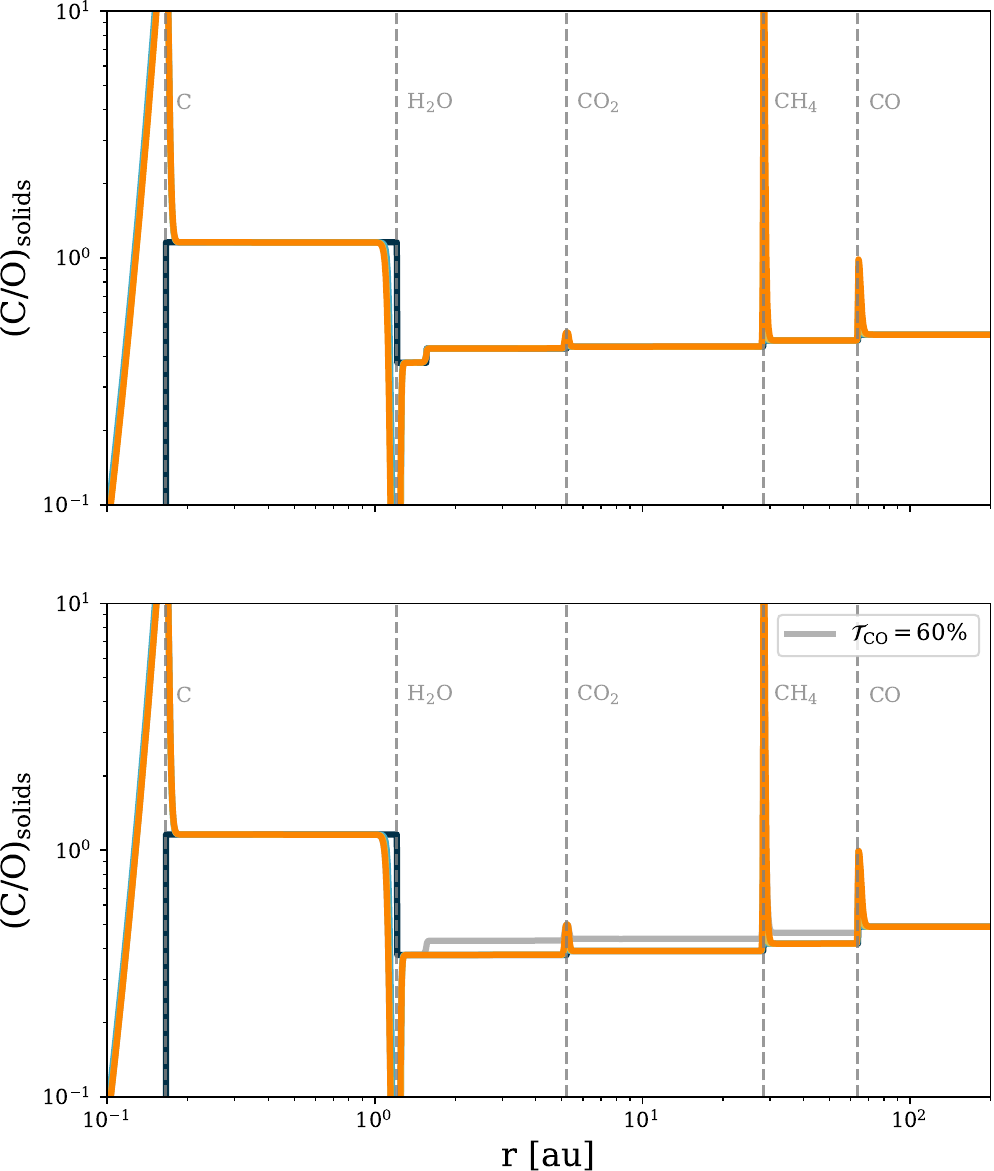}
    \caption{Same as the top half of Fig.~\ref{fig:pebble composition}, showing the temporal evolution of the solid-phase C/O for the \COtrap~=~60\% (top) and no-trapping model (bottom) models. The solid C/O for the \COtrap=~40\% model at $t=5$ Myr is shown as a grey line in the bottom plot for comparison; the trapping model provides a larger solid C/O as dictated by the initial conditions, as discussed in Sec.~\ref{subsec:pebble composition}.}
    \label{fig:solid C/O}
\end{figure}

\section{CO Abundance for $\alpha=10^{-4}$}
\label{appendix:CO abundance}

\begin{figure}
	\includegraphics[width=\columnwidth]{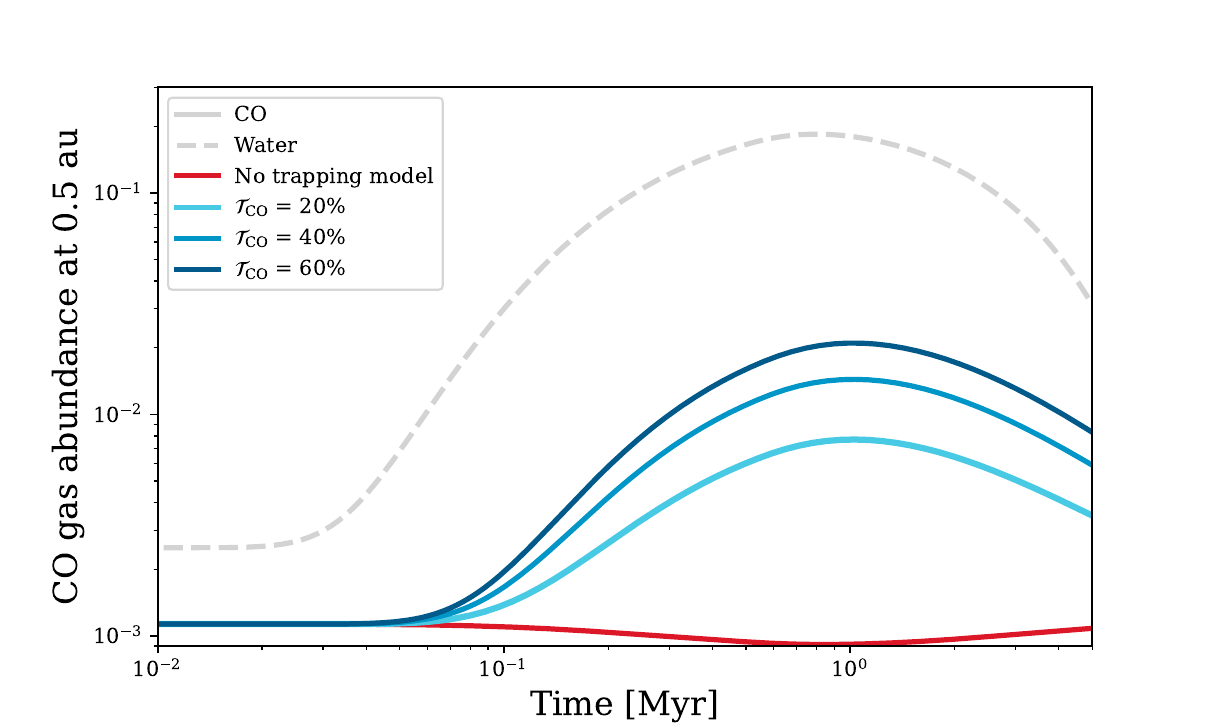}
    \caption{Same as Fig.~\ref{fig:CO abundance}, but for $\alpha=10^{-4}$.}
\end{figure}

\section{Heavy Element Content for $\alpha=10^{-3}$}
\label{appendix:heavy element content}

\begin{figure}
	\includegraphics[width=1\columnwidth]{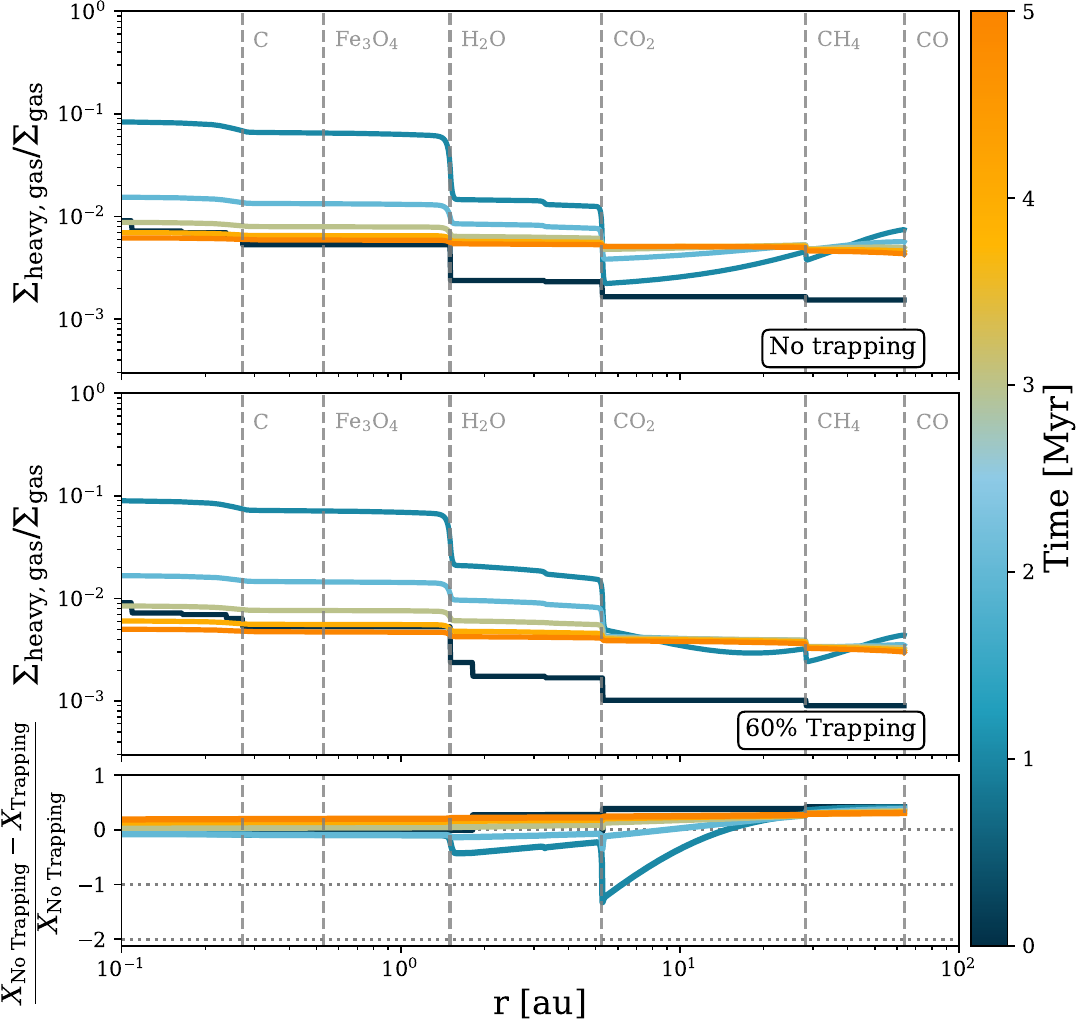}
    \caption{Same as Fig.~\ref{fig:heavy elements}, but for $\alpha=10^{-3}$.}
    \label{fig:heavy elements appendix}
\end{figure}

\bsp
\label{lastpage}
\end{document}